%% file: finalpass5.tex
\documentclass[11pt,preprint]{aastex61}

\usepackage{natbib}
\usepackage {float}
\usepackage{graphicx}
\usepackage{amssymb,amsmath,mathtools}

\received{}
\accepted{}
\shorttitle{Parallax for 7}
\shortauthors{Riess et al.}

\newcommand{\diagentry}[1]{\mathmakebox[1.8em]{#1}}
\newcommand{\xddots}{%
  \raise 5pt \hbox {.}
  \mkern 6mu
  \raise 1pt \hbox {.}
  \mkern 6mu
  \raise -3pt \hbox {.}
}

\newcommand{\kmsmpc}{\hbox{$ \, \rm km\, s^{-1} \, Mpc^{-1}$}}

\newcommand{\bq}{\begin{equation}} 
\newcommand{\eq}{\end{equation}}

\newcommand{\bestscale}{1.037 $\pm$ 0.036 $ \ $}
\newcommand{\bestscalesigma}{3.5}
\newcommand{\bestho}{\hbox{$ 73.48 \pm 1.66 \, \rm km\, s^{-1} \, Mpc^{-1}$}}

\newcommand{\beq}{\begin{equation}}
\newcommand{\eeq}{\end{equation}}
\newcommand{\beqa}{\begin{eqnarray}}
\newcommand{\eeqa}{\end{eqnarray}}

\newcommand{\PL}{$P\hbox{--}L$}

\newcommand{\lcdm}{\hbox{$\Lambda$CDM}}
\newcommand{\tension}{\hbox{$3.7 \sigma$}}

\long\def\check#1{}
\long\def\hide#1{}

\newcommand{\numc}{$ 7 \ $}

\begin{document} 

\title{New Parallaxes of Galactic Cepheids from Spatially Scanning the Hubble Space Telescope: \\ Implications for the Hubble Constant}

\author{Adam G.~Riess}
\affiliation{Space Telescope Science Institute, 3700 San Martin Drive, Baltimore, MD 21218, USA}
\affiliation{Department of Physics and Astronomy, Johns Hopkins University, Baltimore, MD 21218, USA}

\author{Stefano Casertano}
\affiliation{Space Telescope Science Institute, 3700 San Martin Drive, Baltimore, MD 21218, USA}
\affiliation{Department of Physics and Astronomy, Johns Hopkins University, Baltimore, MD 21218, USA}

\author{Wenlong Yuan}
\affiliation{Department of Physics and Astronomy, Johns Hopkins University, Baltimore, MD 21218, USA}
\affiliation{Texas A\&M University, Department of Physics and Astronomy, College Station, TX, USA}

\author{Lucas Macri}
\affiliation{Texas A\& M University, Department of Physics and Astronomy, College Station, TX, USA}

\author{Jay Anderson}
\affiliation{Space Telescope Science Institute, 3700 San Martin Drive, Baltimore, MD 21218, USA}

\author{John W.~MacKenty}
\affiliation{Space Telescope Science Institute, 3700 San Martin Drive, Baltimore, MD 21218, USA}

\author{J.~Bradley Bowers}
\affiliation{Department of Physics and Astronomy, Johns Hopkins University, Baltimore, MD 21218, USA}

\author{Kelsey I.~Clubb}
\affiliation{Department of Astronomy, University of California, Berkeley, CA 94720-3411, USA}

\author{Alexei V.~Filippenko}
\affiliation{Department of Astronomy, University of California, Berkeley, CA 94720-3411, USA}
\affiliation{Miller Senior Fellow, Miller Institute for Basic Research in Science, University of California, Berkeley, USA}

\author{David O. Jones}
\affiliation{Department of Physics and Astronomy, Johns Hopkins University, Baltimore, MD 21218, USA}

\author{Brad E.~Tucker}
\affiliation{Department of Astronomy, University of California, Berkeley, CA 94720-3411, USA}

\begin{abstract} 
We present new measurements of the parallax of \numc long-period ($\geq$ 10 days) Milky Way Cepheid variables (SS CMa, XY Car, VY Car, VX Per, WZ Sgr, X Pup and S Vul) using one-dimensional astrometric measurements from spatial scanning of Wide-Field Camera 3 (WFC3) on the {\it Hubble Space Telescope (HST)}.  The observations were obtained at $\sim 6$ month intervals over 4 years.  The distances are 1.7--3.6 kpc with a mean precision of 45 $\mu$as [signal-to-noise ratio (SNR) $\approx 10$] and a best precision of 29 $\mu$as (SNR = 14).  The accuracy of the parallaxes is demonstrated through independent analyses of $> 100$ reference stars.  This raises to 10 the number of long-period Cepheids with significant parallax measurements, 8 obtained from this program.   We also present high-precision mean $F555W$, $F814W$, and $F160W$ magnitudes of these Cepheids, allowing a direct, zeropoint-independent comparison to $> 1800$ extragalactic Cepheids in the hosts of 19 Type Ia supernovae.  This sample addresses two outstanding systematic uncertainties affecting prior comparisons of Milky Way and extragalactic Cepheids used to calibrate the Hubble constant (H$_0$): their dissimilarity of periods and photometric systems.  
Comparing the new parallaxes to their predicted values derived from reversing the distance ladder gives a ratio (or independent scale for $H_0$) of \bestscale, consistent with no change and inconsistent at the \bestscalesigma $\sigma$ level with a ratio of 0.91 needed to match the value predicted by Planck CMB data in concert with \lcdm.  Using these data instead to augment the Riess et al. (2016) measurement of H$_0$ improves the precision to 2.3\%, yielding \bestho, and the tension with Planck + \lcdm\ increases to \tension.  The future combination of {\it Gaia} parallaxes and {\it HST} spatial scanning photometry of 50 Milky Way Cepheids can support a $< 1$\% calibration of H$_0$.

\end{abstract} 


\section{Introduction} 

   The Hubble constant (H$_0$) measured locally and the sound horizon observed from the cosmic microwave background radiation (CMB) provide the two chief {\it absolute} scales at opposite ends of the visible expansion history of the Universe.   Comparing the two gives a stringent end-to-end test of $\Lambda$CDM (the cosmological constant plus cold dark matter in a flat Universe), the new ``Standard Model'' of cosmology, over the full history of the Universe \citep{Bernal:2016}.  By steadily improving the precision and accuracy of the H$_0$ measurement from Cepheids and Type Ia supernovae (SNe~Ia), evidence has been growing of a significant discrepancy between the two.  The local and direct determination of H$_0$ from \cite[][hereafter R16]{Riess:2016} gives H$_0=73.24 \pm 1.74$ \kmsmpc\ and the most recent value from \cite{Planck-collaboration:2016} in concert with $\Lambda$CDM is $66.93 \pm 0.62$ \kmsmpc, a 3.4$\sigma$ difference.  
   
Intriguingly, this discrepancy does not appear to be attributable to an error in any one source of data, either in the local determination of H$_0$ or from the CMB.
Reanalyses of the R16 data have shown minimal differences in the local determination of H$_0$ with values ranging within $\pm$ 1\%, well within its full 2.4\% uncertainty \citep{Cardona:2017,Follin:2017,Feeney:2017}.  The discrepancy remains significant using any one of three independent, geometric approaches commonly used to calibrate the luminosities of Cepheids along the distance ladder: masers in NGC 4258 \citep{humphreys13}, detached eclipsing binaries (DEBs) in the Large Magellanic Cloud \citep[LMC;][]{Pietrzynski:2013}, or  trigonometric parallaxes of Milky Way (MW) Cepheids \citep{benedict07,vanleeuwen07,riess14,Casertano:2016}.  Replacing Cepheids with the tip of the red giant branch (TRGB) to reach SN~Ia hosts, when possible, produces changes of $< 0.5$\% for the same sources \citep{Jang:2017, Jang:2017a}, as does replacing SN~Ia distance estimates based on optical magnitudes with those relying on the dust-insensitive near-infrared \citep[NIR;][]{Dhawan:2017}.  Indeed, replacing the entire distance ladder with another local Universe distance estimator, time delays from strong gravitational lensing, confirms and reinforces the discrepancy with H$_0 = 72.8 \pm 2.4$ \kmsmpc\ for realistic values of $\Omega_M$ \citep{Bonvin:2017}.

At the early-Universe end of the comparison, \citet{Addison:2017} have shown that the Planck data can be replaced with measurements of $\Omega_B$ via the observed deuterium abundance to calibrate baryon acoustic oscillations (BAO) with no significant change, yielding a predicted value of H$_0=67.0 \pm 1.2$ \kmsmpc\ (but see also \cite{Aylor:2017,DES-Collaboration:2017}).  Yet because this discrepancy may ultimately be interpreted as evidence of new physics in the cosmological model, the burden of proving that it is not the result of a measurement error is necessarily high.  

   Here we present new trigonometric parallax measurements of MW Cepheids using the spatial scanning technique of \citet{riess14} and \citet{Casertano:2016}. 
  These measurements address two outstanding systematic uncertainties associated with prior comparisons of MW and extragalactic Cepheids on the distance ladder: their dissimilarity of periods and photometric systems. 
  The astrometric precision of spatial scanning enables parallax measurements at $D>2$ kpc where most Cepheids with period $P>10$ days live, the period range visible for those in distant SN~Ia hosts.  As we show, scanning is also superior to staring mode for obtaining the precise magnitudes of bright MW Cepheids on the same photometric system as their extragalactic brethren.  
   
  Cepheid variables continue to be the most useful primary distance indicator because they are common, they are sufficiently luminous to be seen with the {\it Hubble Space Telescope (HST)} in the hosts of nearby SNe~Ia ($D \approx 20$--40 Mpc), and they yield an individual distance precision of 0.07 mag \citep{Macri:2015,persson04} when observed in the NIR where they are also insensitive to metallicity \citep{Wielgorski:2017}.  As observed with {\it HST} in the hosts of SNe~Ia at $D \approx 20$--40 Mpc, the reduction in their individual precision ($\sim 0.3$--0.4 mag) due to surface brightness fluctuations in their hosts is offset by typically visible sample sizes of $\sim 100$ per host, providing a mean distance error (statistical) below the uncertainty of the individual SNe~Ia they calibrate ($\sim 0.1$ mag).   
   
   Initially, the use of Cepheids as absolute distance indicators was severely limited by an inability to calibrate their luminosity, once a source of egregious error \citep{Hubble:1929}.  Attaining this calibration demands model-independent, geometric distance measures to the locations of known Cepheids.  Even the nearest known Cepheid (oscillating in the fundamental mode), $\delta$ Cep, has a parallax of only a few mas, requiring sub-mas measurements for useful limits \citep{Gatewood:1993}.  The most precise previous measurements of stellar parallax came from the Fine Guidance Sensor (FGS) on {\it HST}, which can measure relative astrometry to a typical precision of $\sim 0.2$--0.3 mas and thus could usefully measure trigonometric parallaxes to $\sim 10$ known MW Cepheids within 0.5 kpc \citep{benedict07}.  Unfortunately, only two of these have $P \geq 10$ days, the lower end of the period range of those measurable at the typical distances of SN~Ia hosts. Cepheid parallax measurements from {\it Gaia} show tremendous promise and will likely revolutionize such measurements by the mission's final data release in $\sim 2022$.  The first MW Cepheid parallaxes available from {\it Gaia} DR1 have a mean uncertainty of $\sigma \approx 0.3$~mas and parallax SNR $\approx 1$ \citep{Lindegren:2016}.  Although the mean of all $\sim 200$ Cepheids has been used to test and confirm the local measurement of H$_0$, it may not be possible to average so many measurements without incurring additional systematic uncertainties \citep{Casertano:2017}.
   
   The previously mentioned Cepheid calibrations from the LMC and NGC 4258 have specific limitations which continue to motivate the pursuit of MW Cepheid parallaxes.  The LMC is a dwarf galaxy where the environment is quite different from the large spiral hosts used to calibrate SNe~Ia with Cepheids;  its Cepheids are more metal poor by 0.3--0.4 dex \citep{Romaniello:2008}.   As for NGC 4258, its maser distance is unlikely to improve, limiting the ultimate precision available from this anchor to 2.6\% \citep{Riess:2016}.  
   Advancements in the observations of Cepheid trigonometric parallaxes are critical to test the current tension in $H_0$ and to support the future goal of measuring it to 1\% precision.  
   
This is the third paper in a sequence developing and employing a new technique for measuring stellar parallax using spatial scanning observations with the Wide Field Camera 3 (WFC3) on {\it HST}.  We refer the reader to the first \citep[][hereafter Paper I]{riess14} and second applications \citep[][hereafter Paper II]{Casertano:2016}, which provide a thorough description of how spatial scan astrometric data are obtained, calibrated, and analyzed to reach a measurement precision as small as $\sim 30~\mu$as.  Here we provide only a brief overview of the method.

To produce a meaningful measurement (SNR $\approx 10$) of the parallax of $P > 10$ day Cepheids, nearly all of which have $D > 2$ kpc, it is necessary to reach an astrometric precision of $\sim 30~\mu$as for individual epochs of MW Cepheids.  This is factor of 45 better than the ESA astrometric mission {\it Hipparcos} mean for all $P> 10$ day Milky Way Cepheids ($\sigma=1.4$ mas; \cite{vanleeuwen07}), and an order of magnitude better than typically possible with the FGS on {\it HST} or from staring mode observations with WFC3.  For bright sources (MW Cepheids at $D \approx 3$ kpc have $V \approx 9$ mag), spatial scanning\footnote{While shift-and-stare observations using hundreds of exposures could accomplish this as well, readout times and onboard storage limitations make this kind of observing impractical.} provides multiple advantages. In order of decreasing importance, these are (1) a factor of 1000 higher statistical sampling from full detector scans, (2) the ability to average the signal across uncorrected, local distortions along the detector, and (3) removal of jitter along the measurement direction through the comparison to reference-star scans.   These advantages together provide the means to measure parallaxes with 30~$\mu$as precision as described in Papers I and II.   A fourth advantage occurs in the use of Cepheids to measure H$_0$: the ability to measure photometry of MW Cepheids with the same photometric system as those in SN~Ia hosts to cancel zeropoint uncertainties while avoiding additional uncertainties from detector saturation, undersampling and flat field variations.

Here we present a sample of \numc Cepheids, each of which has been observed over 4 years with 8--9 epochs separated by 6 months, double the timespan available in Papers I \& II and for a much larger sample.  In $\S$ 2 we present the spatial scanning measurements and determinations of parallax, and $\S$ 3 gives the spatial scanning {\it HST} photometry of these Cepheids.  We use the prior data in $\S$ 4 to test the scale of the local determination of H$_0$ and to improve the calibration of the MW Cepheid period--luminosity ($P$--$L$) relation.   In $\S$ 5 we discusses the implications and additional considerations.

The HST data used in this paper are available at \dataset[http://dx.doi.org/10.17909/T9QM3G].

\section{Parallax Measurements from Spatial Scans}

The process of extracting high-precision relative astrometry from star trails in spatial scans is described in detail in Papers I and II, and we direct the reader there for further information.  Here we review the key steps discussed in these papers.
  
    $\bullet$  Spatial scans are obtained along the detector $Y$ axis with a $0.05^\circ$ tilt to vary the pixel phase of a point-spread function (PSF) by 1 pixel per 200 pixels scanned.  Astrometric measurements are made along a single dimension, the detector $X$ axis, by fitting a previously determined position-dependent line-spread function (LSF) to each 15 pixel minirow along a star scan, resulting in measurements of the source $X$ position as a function of the $Y$ scan position.
 
    $\bullet$  Astrometric measurements are obtained from deep and shallow scans of the field containing the Cepheid.  The deep scans provide astrometry of fainter reference stars and the shallow scans measure the brighter Cepheids without saturation.   Stars of intermediate brightness ($11 < V < 15$ mag) are used to register the two.
    
    $\bullet$ Pixels struck by cosmic rays and known bad pixels (from prior maps) are excluded from the minirow fits, as are rows whose astrometric measurements in the $X$ direction would be expected to be biased by more than 1 millipixel owing to a nearby star trail (as estimated from their separation and relative brightness).
    
    $\bullet$  Epochs are separated by $\sim 6$ months when the field can be observed by {\it HST} at nominal or $\pm 180\deg$ orientation while maintaining the relative separations of star trails.
    
    $\bullet$ Relative detector coordinates are transformed to relative sky coordinates using a geometric distortion solution following \cite{Bellini:2011} specific to the scan filter.  Additional corrections are made for time-dependent plate-scale variations due to velocity aberration, frame-to-frame rotation, and variable field rotation along the scan.  Lastly, the relative astrometry from independent, coeval scans is registered using a two-dimensional (2D) second order polynomial, $f(x,y)=\Delta X$, to account for simple time-dependent distortions along the measurement direction.
    
     $\bullet$ Relative astrometry along one dimension is measured as the difference from an average, 4-times-oversampled reference line, constructed from the superposition of all time-aligned scan lines.  The reference line contains the jitter history, length, and slope of the scan.
     
     $\bullet$ An additional static perturbation to the geometric distortion solution from \cite {Bellini:2011} was computed for three filters ($F606W$, $F621M$, and $F673N$) based on repeated scans of stars in M48 and M67 (GO 14394), and applied as described in Paper II. 
     
     $\bullet$ Repeated scans demonstrate measurement precision (i.e., residuals) for well-sampled bright stars at the 0.5--2 millipixel level (20--80 $\mu$as), with the higher precision obtained for scans with similar conditions (orient, position along breathing-focus cycle) and lesser precision for scans with orients differing by $180^\circ$.
  
     $\bullet$ A multisource, multiparameter model is fit simultaneously to the Cepheid and the reference stars in the field using as additional constraints the spectrophotometric (hereafter SP) parallax priors for the reference stars.  These priors are derived from the comparison of stellar magnitudes from {\it HST} in typically two UV filters ($F275W$, $F336W$), four Str\"{o}mgren medium bands ($F410M$, $F467M$, $F547M$, $F621M$), two NIR bands ($F850LP$, $F160W$), as well as catalog 2MASS $JHK$ and {\it WISE} Band-1 and Band-2 magnitudes.  In addition, the stellar model for a reference star is constrained with spectra which are used to determine the MK spectral class and luminosity class of each star \citep{Gray:2014}, which constrains their temperature and specific gravity.  A best-fitting stellar model for each star is determined from the Padova Isocrones \citep{Bressan:2012} dependent on three physical parameters (temperature, log gravity, metallicity) and two environmental parameters (extinction and distance).  A line of sight, one-dimensional (1D) extinction probability prior for stars along the line of sight based on 2MASS data \citep{Marshall:2006} augments the fits, as does a model of the frequency of the stellar types along each line of sight for the Milky Way \citep{Robin:2003}.  The reference stars in each field are found to have a mean distance of $\mu=11.7$--12.0~mag with a mean individual uncertainty of $\sim 0.2$ mag.  The typical error in the reduction from relative to absolute parallax is $\sim 10~\mu$as (Paper II) and is found to be subdominant to the astrometric measurement errors.
     
     $\bullet$ each star's 1D motion is modeled as the superposition of a relative proper motion and parallax.  Stars whose 8 to 9 epochs (4~yr span) poorly fit this model (e.g., astrometric binaries), typically $\sim 10$\%--20\% of all stars, are iteratively rejected until the fit converges.  Extensive observations of high-precision radial velocities were used by \citet{Anderson:2016} to rule out significant contamination of the parallax measurements (an upper limit of $< 4$\%) due to binarity for the Cepheids presented here.
     
     $\bullet$ The global fit makes use of the parallax factors or the projection of the parallax ellipse on the $X$-axis of the detector at the measurement epoch.  By definition, the parallax factors are a number between $-1.0$ and 1.0, and in practice they have a mean absolute value of 0.8.  

     In Table 1 we list the observations for each Cepheid field.  Figures 1 to 7 show the reference stars and Cepheid in each field. Table 2 lists the parameters and characteristics of the reference stars in each of the Cepheid fields. 
    
     As in Paper II, we test the fidelity of our astrometric parallax determinations by comparing them to the SP parallax priors for the reference stars.  To make this comparison independent of the SP measurement, we remove each from the set of prior constraints before recomputing the best global fit.  The resulting astrometric parallax is then compared to its independent SP parallax. These are shown for all of the reference stars in Figure 8.  On average the agreement is quite good.  We note that the astrometric parallax measurement for the Cepheid has the advantage of using all of the reference-star SP priors or one more than these tests for the reference stars.
     
     In Figure 9 we display the proper-motion-subtracted astrometric measurements of the Cepheids and their best-fit parallaxes.   We give the Cepheid parallaxes and their uncertainties in Table 3, showing the contribution from the reduction from relative to absolute parallax which is subdominant in all cases.  Differences in the Cepheid parallax uncertainties are largely produced by the differing availability of uncontaminated reference stars used to register deep and shallow scans as discussed further in \S 5.\footnote{The process of astrometrically registering shallow and deep scans while retaining a precision of a few millipixels necessitates the determination of 7 free parameters including an offset, frame rotation, and 5 polynomial parameters to solve for second-order in $x$ (time-dependent geometric distortions likely due to the thermal cycle of the telescope).  Thus, a robust solution requires more than 7 reference stars with high SNR in both scans, not saturated in the deep scan, running most of the scan length, and not crossing another star trail.  Unfortunately, a number of the Cepheids initially targeted for spatial scanning do not have this number of reference stars and their monitoring was curtailed after 5 epochs.  In the future, an ability to predict the distortion produced by the thermal cycle could reduce the number of necessary stars to $\sim 2$--3 and allow a complete analysis of all observed Cepheids with improved relative astrometry.}

  \startlongtable
\begin{deluxetable*}{cccc}
\tabletypesize{\small}
\tablewidth{0pt}
\tablenum{3}
\tablecaption{Parallax}
\tablehead{\colhead{Cepheid} & \colhead{$\pi$ (milliarcsec)} & \colhead{$\sigma_{\rm tot}$} & \colhead{$\sigma_{\rm abs}$} }
\startdata
SS CMa & 0.389 & 0.0287  & 0.003   \\
XY Car & 0.438 & 0.0469  & 0.011   \\
VX Per & 0.420 & 0.0744  & 0.017   \\
VY Car & 0.586 & 0.0438  & 0.009   \\
WZ Sgr & 0.512 & 0.0373  & 0.011   \\
S Vul  & 0.322 & 0.0396  & 0.007    \\
X Pup  & 0.277 & 0.0469  & 0.009    \\
\hline
\enddata
\tablecomments{$\sigma_{\rm abs}$ is the contribution to $\sigma_{\rm tot}$ from the reduction from relative to absolute parallax.}
\end{deluxetable*}

    \section{Rapid Scan WFC3 System Photometry}
    
    The best way to reduce the propagation of zeropoint and bandpass uncertainties among Cepheid flux measurements along the distance ladder is to observe all Cepheids with a single, stable photometric system.  This is especially important in the NIR where individual system zeropoint uncertainties are $\sim 0.02$--0.03 mag \citep{riess11c} and the relative differences between two systems is thus $\sim 0.03$--0.04 mag.  This is challenging to accomplish for MW Cepheids which are 14--18 mag brighter than their extragalactic counterparts.
     
     Spatial scanning offers several of advantages to achieve this goal.  Scanning varies the position of the source on the detector, which averages down errors in the flat fields and can also be used to vary the pixel phase, reducing the uncertainty from undersampled PSF photometry.  By scanning at high speeds, it is possible to reliably reduce the {\it effective} exposure time far below what is possible in staring mode owing to variations and uncertainties in shutter flight time \citep{Sahu:2015}.

      To observe these MW Cepheids in the range of $7<V<10$ mag, we utilized the highest available scan speeds of $7.5''$~s$^{-1}$ (under gyro control) in $F160W$ to realize an effective exposure time of $\sim 0.01$~s.  In $F555W$ and $F814W$ scan speeds were either $7.5''$ s$^{-1}$ or $4.0''$ s$^{-1}$.
    
    Photometry is measured from the amplitude of the fit 
of the LSF to the extracted signal in each 15-pixel 
minirow — the same fit used for the astrometric 
measurement of the source’s position along the $X$ direction -
divided by the effective exposure time (i.e., the pixel size
divided by the scan rate). Empirical measurements of the scan rate show very good agreement with the commanded rate, with the latter
preferred as it is less noisy.  However, as we will show, we use pairs of direct and scanning images to calibrate out a possible error in the pixel size and scan rate.

Because of the variable geometric distortion, two 
corrections must be applied to convert this amplitude 
into a uniform flux measurement across the detector, 
unaffected by the local pixel size.  The first is to
multiply the measured flux by the local (relative) 
pixel area using the same pixel area map (or distortion solution to rectify images) used for photometry of all point sources in staring mode.\footnote{A pixel area map is used to normalize the flux for  photometry to compensate for the fact that the 
flat field for WFC3 images is defined so that a uniform 
illumination produces equal counts per pixel, regardless 
of their size.  Therefore the local measured count 
rate is actually proportional to the flux per unit area on the sky.}
The second is a correction for the different size of each 
pixel along the scan ($Y$) direction: larger pixels take 
longer to be traversed, and thus experience a longer 
effective exposure time.  The second correction 
partially compensates the first, so that no correction 
is actually needed for the pixel size along $Y$,
and the net correction needed is to multiply 
the fitted amplitude by the relative pixel size in 
the $X$ direction.
    
    By comparing pairs of scans of MW Cepheids (programs GO-13335, GO-13928) in back-to-back exposures, we find a mean photometric error per scan observation of 0.007, 0.003, and 0.001 mag in $F160W$, $F555W$, and $F814W$, respectively.  
    The onset of saturation at $7.5''$ s$^{-1}$ occurs for stars brighter than 7.2, 6.8, and 5.6 (Vega) mag in $F160W$, $F555W$, and $F814W$, respectively. 
       Within 1~mag (brighter) of this saturation limit, the precision of spatial scan photometry is seen to decrease to 0.01--0.02 mag in $F555W$ and 0.003--0.01 mag in $F814W$.  The precision in F160W does not significantly decrease within 2 mag (brighter) of the onset of saturation.  In general, photometry of mildly saturated Cepheids (i.e., loss of the central 2--3 pixels) remains robust and well constrained in these scans (also seen in comparison to ground-based observations) because of good knowledge of the WFC3 PSF, the stability of WFC3, the natural averaging of measurements over position on the detector, and the variation in source pixel phase which mitigates undersampling of the PSF.   
            
       To insure uniformity between photometry in imaging modes, we use a large set of images of reference stars obtained in pairs, one staring and one scanning, to find and measure the offset between the two.  We expect a constant-magnitude offset as it includes the aperture correction of the scanned data LSF from its 15-pixel minirow extent to an infinite aperture as used for imaging data.  In addition, measuring this empirically would include and thus cancel an error in the combination of the pixel size and scan rate (used to define the effective exposure time), as these same values are used for the reference-star scans.  The same programs (GO-13335, GO-13928) used to obtain additional Cepheid scans also obtained staring mode images and contain additional reference stars in a useful magnitude range.  In $F160W$,  there were 88 observations of stars in the fields of the Cepheids in a useful magnitude range of $12.5 < F160W < 14.0$.  The difference between the scanned and staring photometry in $F160W$ has a dispersion of 0.03 mag, primarily from staring mode errors due to undersampling and flat field variation, and the offset has an error in the mean of 0.0033 mag.  The same comparison at a much slower speed ($0.31''$ s$^{-1}$ instead of $7.5''$ s$^{-1}$) for 10 stars in M35 yields a consistent result (GO-13101).  In $F555W$ and $F814W$, 17 stars in M67 (GO-14394) in the range 11--15 mag were scanned ($0.40’’$ s$^{-1}$) and imaged, and used to derive a constant offset between the scanning and staring mode apertures with a precision of 0.003 and 0.002 mag, respectively.  This was successfully tested against 11 stars in the frames of the Cepheids at the higher scan speed.  

    When using Cepheid magnitudes to measure distances, it is useful to make use of a ''phase correction'', the difference between the Cepheid magnitude at the observed phase and at the (flux) mean of their light curve, since the magnitude (at the time of mean flux) is the best distance indicator for Cepheids.  These phase corrections are most efficiently derived from  ground-based light curves of these Cepheids in matching filters. Because the phase corrections are relative quantities, they do not change the zeropoint of the light curves, which remain on the {\it HST} WFC3 natural system.  The phase-correction uncertainties are determined from the variations around a locally smooth curve. The algorithm for determining the phase corrections and the sources of the ground-based data are given in the Appendix.
   
       The ground-based Cepheid light curves are shown in Figure 12 together with the {\it HST} photometry.  In $F555W$ and $F814W$ each Cepheid was observed at 3--5 epochs.  In $F160W$ each was observed at 3--6 epochs.
    
    With photometry of this precision, the use of individual epochs to calibrate the luminosity of Cepheids is dominated by the uncertainties in the phase corrections.  The mean uncertainties in these phase corrections per epoch are 0.035, 0.023, and 0.018 mag in $F555W$, $F814W$, and $F160W$ (respectively).  The phase-correction errors scale with the size of magnitude changes in the light curves and are thus smallest in the NIR. These values are consistent with the dispersion of multiple epochs measured for the {\it HST} observations around the ground-based light curves.  The mean uncertainty in the light-curve mean magnitude for these 7 Cepheids is 0.021, 0.010, and 0.009 mag in $F555W$, $F814W$, and $F160W$, respectively.  At this level of precision, parallax uncertainties will dominate the determination of absolute luminosities for uncertainties greater than 5 $\mu$as. 
    
     In Table 4 we provide the photometric measurements of these \numc Cepheids for WFC3 $F555W$, $F814W$, and $F160W$.

     \section{A Test of the Calibration of the Hubble Constant}
     
     Here we seek to test the Cepheid calibration of the Hubble constant.  
     However, for simplicity, we will initially avoid contending with the well-known bias that arises in the conversion of parallax measurements with modest SNR to absolute magnitudes, often referred to as the Lutz-Kelker-Hanson bias \citep[LKH;][]{Hanson:1979}. Because the uncertainty in the parallaxes we would {\it predict} for these Cepheids based on their periods and magnitudes is a factor of 15--20 smaller than their measured parallax uncertainties, we compare the unbiased parallax predictions (expected bias $< 0.001$ mag) directly to the measurements. Carrying out this fit in parallax space avoids nonlinear 
   transformations of the relatively large parallax error.  
   A similar method has been used, e.g., by \cite{feast97} in their luminosity calibration of Cepheids   from Hipparcos parallaxes; see \cite{Sesar:2017} for a recent
   detailed discussion.
 We will make use of the measured parallaxes to determine absolute magnitudes and account for the LKH bias in \S 4.1.  
     
      Using the known periods of these Cepheids and their photometry in Table 4, we can {\it predict} their parallaxes using the distance-ladder parameters presented by \citet{Riess:2016} as in \cite{Casertano:2017}.  This is equivalent to reversing the distance ladder to predict the parallaxes from $H_0$.  First, we form the same Wesenheit reddening-free magnitudes used by R16 \citep{madore82}: 
    
    \bq m^W_H=m_{F160W}-0.386(m_{F555W}-m_{F814W}). \eq
\noindent
    These $m^W_H$ values have a mean uncertainty of 0.012 mag.

    There is a small count-rate nonlinearity effect (hereafter CRNL) which can make faint sources appear too faint in HgCdTe devices like WFC3-IR due to charge trapping.  This term has been measured to be $0.008 \pm 0.003$ mag dex$^{-1}$ for WFC3-IR \citep{riess11b, riess18}.  We correct for the CRNL by adding $0.026 \pm 0.009$ mag to account for the 3.2 dex between the observed count rates of MW Cepheids and the extragalactic Cepheids of \citet{Riess:2016}. 

The best-fit solution from \citet{Riess:2016} yields a value of H$_0$ of 
H$_0 = 73.24$ \kmsmpc.  Based on this value, the derived calibration of the Cepheid $P$--$L$ relation is

\bq M^W_H=-5.93-3.26({\rm log} P - 1).  \eq 
\noindent
Employing the derived periods (see Appendix) yields the values of $M^W_H$, and combined with the apparent Wesenheit magnitudes on the WFC3 system ($m^W_H$)

we derive distance moduli of \bq \mu=m^W_H - M^W_H \eq
\noindent
and the expected parallax

\bq \pi_{R16}=10^{-0.2(\mu-10)} \eq in mas.      
With negligible uncertainties in the periods, the mean uncertainties in the predicted $\mu$ are 0.015 mag, or $< 1$\% in distance.  These expected parallaxes on the scale in which H$_0=73.24$ \kmsmpc\
are given in Table 4 as $\pi_{R16}$.  (These are weakly correlated owing to the CRNL correction uncertainty of 0.009 mag, but this is insignificant compared to the measured uncertainties in the parallaxes.)

In Figure 10 we compare the predicted and measured parallaxes for the 8 Cepheids from this program (7 presented here and SY Aur from Paper I: measured = $0.428 \pm 0.054$ mas, predicted = 0.418 mas).  The agreement between the measured and predicted parallaxes is reasonably good: total $\chi^2 = 12.9$ for 8 points, with a higher value expected to occur 15\% of the time by chance.

Now we can consider whether there is a better scale for the predicted parallaxes --- that is, a preferred value of H$_0$ that improves the agreement with the measured parallaxes.  Such a value of $\alpha$ would favor a new value of $H_0$ from an independent calibration of the Cepheids as $H_{0,new}=\alpha (H_{0,R16})$. To determine this, we minimize the usual $\chi^2$ statistic for the set of 8 parallaxes,

\bq \chi^2=\sum {\alpha\pi_{R16}-\pi_{obs} \over \sigma}, \eq with respect to a free parameter ($\alpha$) used to vary the distance scale of R16, that is $\pi_{obs}=\alpha \pi_{R16}$.
\noindent
We find a value of $\alpha$=\bestscale\ (with the best $\chi^2 = 11.9$), as shown in Figure 11, consistent with no rescaling.  

Reducing the predicted parallaxes by 9\%, equivalent to increasing their mean $\mu$ by 0.20 mag ($\alpha=0.91$) to place them on the H$_0=67$ \kmsmpc\ scale of Planck \citep{Planck-collaboration:2016} and \lcdm, results in a total $\chi^2=24.2$, or an increase of 12.3 for one degree of freedom, a result disfavored at the \bestscalesigma $\sigma$ (99.9\%) confidence level. This result is independent of the use of the masers in NGC 4258, the DEBs in the LMC, and the parallaxes of MW Cepheids measured with the {\it HST} FGS from \citet{benedict07} and augmented by {\it Hipparcos} \citep{vanleeuwen07}, used by R16 to determine H$_0$.  
    
    \subsection{The Milky Way Cepheid $P$--$L$ Relation and $H_0$}
    
    Use of the measured Cepheid parallaxes with modest SNR and their apparent magnitudes to estimate their absolute magnitudes requires a small correction for the LKH bias, which we estimate following \citet{benedict07} and \citet{vanleeuwen07} (Table 4).\footnote{
    This bias does not apply when comparing 
   different measurements of the parallax for
   the same object, and is negligible for the conversion of high-SNR measurements of magnitudes to parallax.}

        This bias may be considered in two parts: (a) the change in shape of the PDF of the parallax estimate when used to estimate its inverse, i.e., $D=\pi^{-1}$ or $\mu=5{\rm log}D+10$, and (b) the nonuniformity of the {\it a priori} distribution of possible distances (and ultimately magnitudes) for a disk population of stars with a given measured parallax, i.e., there are more stars with $D=\pi^{-1}+\sigma$ consistent with a parallax measurement than those with $D=\pi^{-1}-\sigma$.  For each parallax measurement, we use the density distribution used by the Besan\c{c}on Model of the Galaxy \citep{Robin:2003} for young ($< 1$ Gyr) stars
        to produce a realistic, distribution of stars along each line of sight.  We then average their magnitudes weighted by their consistency with the parallax measurement and derive the LKH bias as the difference of this average with the value derived from the nominal measured parallax.  Our estimates of the LKH bias match those of \citet{vanleeuwen07}, with a mean difference of $< 0.001$ mag and an individual dispersion of 0.02 mag, and are listed for the new parallaxes in Table 4.
    
    The debiased distance modulus is then given by 
    
    \bq \mu=5 {\rm log} \pi_{\rm obs}^{-1} + 10 - LKH \eq
\noindent
and the absolute magnitude is derived from $\mu$ and the flux (see Equations 1 and 3).  The absolute $P$--$L$ relation in the WFC3 Wesenheit system for these 8 MW Cepheids is shown in Figure 13 together with the results from \citet{benedict07} and \citet{vanleeuwen07}.  
The new set is consistent with the prior measurements, but it now usefully extends the well-sampled range to $P>10$ days while avoiding the uncertainties in transforming from ground-based to the WFC3 photometric system.   
    
    Following R16, we can utilize this expanded \PL $ \ $ relation to help calibrate the luminosity of SNe~Ia and improve the determination of the value of H$_0$.  Of the three geometric sources of Cepheid luminosity calibration used by R16 to determine H$_0$ (masers in NGC 4258, DEBs in the LMC, and MW Cepheid parallaxes), the MW Cepheids yielded the highest value of $76.2 \pm 2.4$ \kmsmpc, which is 1.5$\sigma$ higher than the mean of the other two, $71.6 \pm 1.8$ \kmsmpc.  It is not uncommon for the highest measurement among three to differ at this level from the mean of the lower two (probability of 17\%).  As noted in the previous section, the new parallax measurements modestly decrease the inferred value of H$_0$ when used with the prior MW Cepheid parallaxes and yields $75.9 \pm 2.1$ \kmsmpc.  Including the new parallax measurements with all three anchors yields H$_0=73.48 \pm 1.66$ \kmsmpc, a reduction in total error (including systematics) from 2.37\% to 2.27\%.  There is a small {\it increase} in H$_0$ of 0.24 \kmsmpc\ from the R16 value owing to a small covariance between the parameters and measurements of the other two anchors.  Comparing the difference between this locally determined value of H$_0$ and the value from Planck and \lcdm \citep{Planck-collaboration:2016} increases the significance of the difference from 3.4$\sigma$ seen by R16 to now 3.7$\sigma$.

\startlongtable
\begin{deluxetable*}{cccccccccccccccc}
\tabletypesize{\scriptsize}
\tablewidth{0pt}
\tablenum{4}
\tablecaption{Photometric Data for MW Cepheids\label{tb:phot}}
\tablehead{\colhead{Cepheid} & \colhead{$F555W$} & \colhead{$\sigma$} & \colhead{std } & \colhead{eps} & \colhead{$F814W$} & \colhead{$\sigma$} & \colhead{std} & \colhead{eps} & \colhead{$F160W^a$} & \colhead{$\sigma$} & \colhead{std} & \colhead{eps} & \colhead{Period} & \colhead{$\pi_{R16}$} & \colhead{LKH}}

\startdata
S VUL & 9.138 & 0.028 & 0.050 & 3 & 6.856   & 0.010 & 0.017 & 3 & 4.885   & 0.009 & 0.026 & 4 &   68.966   & 0.289 & -0.12 \\
SS CMA & 10.133 & 0.023 & 0.035 & 3 & 8.440 & 0.013 & 0.005 &  4 & 6.892   & 0.011 & 0.023 &  3 &  12.356   & 0.317 & -0.03 \\
VX PER & 9.460  & 0.010 & 0.011 &  4 & 7.913 & 0.006 & 0.017 & 3 & 6.471   & 0.009 & 0.021 & 5 &  10.887   & 0.407 & -0.10 \\
VY CAR & 7.591  & 0.029 & 0.102 & 5 & 6.224 & 0.008 & 0.040 & 5 & 4.972   & 0.007 & 0.031 & 6 &   18.898   & 0.549 & -0.06 \\
WZ SGR & 8.177  & 0.016 & 0.027 & 5 & 6.476 & 0.012 & 0.034 & 5 & 4.856   & 0.010  & 0.046 &  4 &  21.851   & 0.559 & -0.06 \\
X PUP & 8.692  & 0.026 & 0.016 & 3 & 7.129 & 0.010 & 0.015 & 3 & 5.626   & 0.008  & 0.020 & 4 &   25.967   & 0.342 & -0.10 \\
XY CAR & 9.466 & 0.014 & 0.027 & 4 & 7.926 & 0.010 & 0.016 & 3 & 6.456   & 0.006  & 0.019 & 6 &  12.436   & 0.376 & -0.07 \\
\hline
\enddata
\tablecomments{$^a$Does not include addition of $0.026 \pm 0.009$ mag to correct CRNL between MW and extragalactic Cepheids.}
\end{deluxetable*}

\section{Discussion}

The 8 MW Cepheid parallax and photometric measurements presented here provide an alternative source to calibrate the Cepheid \PL $ \ $ relation with two advantages over the prior MW sample: (1) their periods, with a range of $10 < P < 69$ days and a mean of 18 days, provide a better match to the $P>10$ day extragalactic Cepheids detectable in SN~Ia host galaxies, and (2) they use the same WFC3 photometric system as these extragalactic Cepheids.  As previously seen with all three anchors used by R16, the new data confirm the tension seen with \citet{Planck-collaboration:2016} used in conjunction with $\Lambda$CDM to predict the value of H$_0$.\footnote{\citet{Anderson:2017} have estimated a small bias (0.2\%) in the value of H$_0$ owing to the association of Cepheids with clusters and the changing resolution along the distance ladder.  Correcting for this association bias would decrease H$_0$ to 73.27 \kmsmpc\ and lower the tension with Planck from 3.7 to 3.6$\sigma$.}    With now four independent sources of calibration, it is very hard to imagine that systematic errors in the calibration of the distance ladder are the root cause of the tension. 
As discussed in \S 1, neither a replacement of the Cepheids as the bridge to SNe~Ia nor of the SN~Ia optical magnitudes produces a meaningful shift in the local determination of the Hubble constant.  Elsewhere it has been shown that variations between the local and global value of H$_0$ are expected and observed to be $\sim 0.3$\% \citep{Odderskov:2014, Odderskov:2017, Riess:2016, Wu:2017}, a factor of 30 smaller than the present 9\% difference.

Additional precision in the calibration of the distance ladder is expected from forthcoming releases of the ESA {\it Gaia} mission.  Already Data Release 1 (DR1) provided hundreds of Cepheid parallax measurements.  Although the DR1 parallaxes had an order of magnitude less precision than those presented here (see Figure 10), they are more numerous, and the precision by the expected final data release in $\sim$2022 is expected to be a factor of 4--5 {\it better} than what we have achieved with the WFC3 scanning approach.  While it is not yet clear if one can reasonably combine hundreds of low-SNR measurements of DR1 Cepheid to produce one high-SNR measurement without penalty, \citet{Casertano:2017} showed that the result of doing so was a scale of $\alpha=0.997 \pm 0.031$, consistent with the result of \bestscale from the set of 8 independent parallaxes presented here.  

To make optimal use of the parallaxes from {\it Gaia}, with individual uncertainties expected to reach $\sim 3$\% for hundreds of Cepheids, it is even more important to measure their photometry on the same {\it HST} WFC3 system as their extragalactic cousins.  Failure to do so would leave the system-to-system uncertainty in the NIR 
of $\sigma \sim 0.03$ mag \citep{Skrutskie:2006, Kalirai:2009}, precluding a 1\% determination of H$_0$.  The match between the ground-based $H$-band and the WFC3 $F160W$ is particularly poor as indicated by the large color term of $\sim 0.2$ mag per mag of $J-H$ color which has been measured between the two \citep{Riess:2016, riess11c}.

To enable this goal, we have collected $F555W$, $F814W$, and $F160W$ photometry of 50 of the most useful MW Cepheids ($P>10$ days, lowest extinction, nearest, brightnesses suitable for {\it Gaia}) using the same approach as in \S 3.  
The expected error for this sample of 50 would be under 0.4\% each for their mean parallax and photometry, and would keep the total uncertainty in $H_0$ to under 0.6\%, a suitable anchor for a 1\% determination of H$_0$.  Assuming comparable improvement in the tie between Cepheids and SNe~Ia (50-60 systems), a measurement with this precision would match the precision of the CMB-based prediction and provide a powerful end-to-end test of the cosmological model.

\bigskip

\bigskip

\acknowledgements

This project was enabled by significant assistance from a wide variety of sources. Gautam Upadhya, Aviv R.~Cukierman, Merle Reinhardt, George Chapman, William Januszewski, and Ken Sembach provided help with the {\it HST} observations. We thank Ryan Foley and Yen-Chen Pan for obtaining spectra of two reference stars with SOAR and Peter Challis for taking a few spectra of stars that were not ultimately used here.  We thank Ed Nelan, Matt Lallo, Fritz Benedict, and Barbara McArthur for productive discussions about the behavior of the FGS. We also thank Leo Girardi, Alessandro Bressan, and Paola Marigo for the use of and assistance with their Padova isochrone database, Anne Robin for assistance with the Besancon Galaxy Model, Eddie Schlafly and D. Marshall for input on the extinction along the line of sight to these stars, and Nolan Walborn for useful discussions on the classification of hot stars.  We thank Richard Anderson for useful discussions.  We thank Scott Engle for observations relating to Cepheid period observations and Richard Gray for help with MK classification.  We thank Kris Stanek for additional observations from the ASAS.

Support for this work was provided by the National Aeronautics and Space Administration (NASA) through programs GO-12879, 13334, 13335, 13344, 13571, 13678, 13686, 13928, 13929, 14062, 14394, 14648, 14868 from the Space Telescope Science Institute (STScI), which is operated by AURA, Inc., under NASA contract NAS 5-26555. A.V.F.’s group at UC Berkeley is also grateful for financial assistance from NSF grant AST-1211916, the TABASGO Foundation, the Christopher R. Redlich Fund, and the Miller Institute for Basic Research in Science (UC Berkeley).
 S.C. and A.G.R. gratefully acknowledge support by the Munich Institute for Astro- and Particle Physics (MIAPP) of the DFG cluster of excellence ``Origin and Structure of the Universe.'' Research at Lick Observatory is partially supported by a generous gift from Google.

This research is based primmarily on observations with the NASA/ESA {\it Hubble Space Telescope}, obtained at STScI, which is operated by AURA, Inc., under NASA contract NAS 5-26555. Some of the data presented herein were obtained at the W. M. Keck Observatory, which is operated as a scientific partnership among the California Institute of Technology, the University of California, and NASA;
the Observatory was made possible by the generous financial support of the W. M. Keck Foundation. This publication makes use of data products from the {\it Wide-field Infrared Survey Explorer (WISE)}, which is a joint project of the University of California (Los Angeles) and the Jet Propulsion Laboratory/California Institute of Technology, funded by NASA. It has also made use of the SIMBAD database, operated at CDS, Strasbourg, France.
   
 %\section{Appendix: Phase Corrections}  

  \appendix
   
     We made use of long-span multiband photometric measurements from the ground for each Cepheid in four bands ($V$, $I$, $J$, and $H$) to determine the phase corrections (the magnitude difference between an observed phase and the magnitude at the flux mean of the light curve) in the three similar WFC3 bands ($F555W$, $F814W$, and $F160W$).  
     
 We based our phase determination on seven datasets: (1) $V$-band measurements from the ASAS catalog of variable stars \citep{Pojmanski1997}; (2) $V$-band measurements from the INTEGRAL Optical Monitoring Camera \citep{Gimenez:2001}; (3) multiband literature photometry measurements from the McMaster Cepheid Photometry and Radial Velocity Data Archive; (4) $I$-band measurements from \citet{Berdnikov+2000}; (5) $JHK$ measurements from \citet{Monson+2011};  (6) our own $H$-band measurements using the CTIO 1.3~m telescope obtained in 2014--2015, and (7) $V$-band measurements from the 1.3 meter RCT at Kitt Peak (Scott Engle, private communication).  Data from different sources but with similar filters were merged to the same bands. For each band, we shifted light curves of various sources to a common zeropoint and rejected statistical outliers prior to the phase determination. The data used for each Cepheid are given in Table 6.

We fit the combined dataset with Fourier series in order to create a model from which to estimate the phase correction. For each Cepheid, the combined measurements can be described by an $n\times 4$ matrix $\{t, m, \sigma, b\}$, where $n$ is the number of measurements, $t$ is the observation time, $m$ is the magnitude, $\sigma$ is the photometric uncertainty, and $b$ is a band label. We ordered the dataset by $b$ and shifted $t$ such that the midrange of the observation dates is zero. For each band we construct a $n_b\times(2k_b+1)$ matrix,
\begin{equation}
\begin{split}
B_b = \{&1, \cos(\omega t_{bi}), \sin(\omega t_{bi}), \cos(2\omega t_{bi}), \\
&\sin(2\omega t_{bi}), ... , \sin(k_b\omega t_{bi})\}, \\
i = 1, 2&, ..., n_b,
\end{split}
\end{equation}
\noindent
where $n_b$ is the number of measurements in band $b$, $t_{bi}$ is the time of the $i$th measurement in band $b$, $k_b$ is the order of Fourier series for band $b$, and $\omega=2\pi/P$ for a Cepheid with constant period $P$. Since the light-curve shapes can be different across different bands while the Cepheid pulsating phase does not depend on wavelength, we construct the design matrix as 
\begin{equation}
X = 
\begin{pmatrix}
\diagentry{B_1}\\
&\diagentry{B_2}\\
&&\diagentry{\xddots}\\
&&&\diagentry{B_M},\\
\end{pmatrix}
\end{equation}
\noindent
where $M$ is the total number of bands. All of the off-diagonal blocks are filled with zeros. The diagonal blocks share the same $\omega$ while their Fourier coefficients can be different. We perform a grid search on the free parameters in $\omega$ and adopt the fit with the least error-weighted residual sum of squares --- i.e., the total $\chi^2$ of the fit to the data. Given a trial $\omega$, the regression coefficients are solved by 
\begin{equation}
\beta = (X^T\Sigma^{-1}X)^{-1}\, (X^T\Sigma^{-1}Y),
\end{equation}
\noindent
where $\beta$ is the best-fit array of Fourier coefficients, 
$\Sigma$ is a rank $n$ diagonal matrix with $\mathrm{Diag}(\Sigma) = \{\sigma^2\}_n$, and $Y = \{m\}_n$ is an $n\times 1$ matrix for the measured magnitudes. The $\chi^2$ is subsequently obtained by 
\begin{equation}
\chi^2 = (Y-X\beta)^T\Sigma^{-1}(Y-X\beta).
\end{equation}

We tried two models to describe $\omega$. The first is a constant-period model with $\omega = 2\pi/p_0$, where $p_0$ is the period $P$, while the second is a varying-period model with $\omega = 2\pi/(p_0 + p_1\, t)$, where $p_0$ and $p_1$ are free parameters. Because these Cepheids are very bright, the values of $\sigma$ are low and fairly equivalent, so in practice we weight all of the data equally.
 We excluded any bands with fewer than 20 measurements, full spans less than a year, or with phase gaps larger than 0.15 from phase determination. The order of the Fourier series for each band was set to $\sim 1/4$ of the number of measurements $n_b$, and we restricted the order to be no higher than 7. We found that the period change of S~Vul is sufficiently complicated over the time span of collected data that its phase cannot be described by the above simple models. To obtain a better estimation of its phase, we excluded any data older than HJD = 2,454,500 and fit the model only with the data close in time to the {\it HST} observations. We adopted the varying-period model for all Cepheids except WZ Sgr.
 The end result of the Fourier model is the Cepheid period or period function.  Because light curves in all bands are useful for determining the period or its change, we used a broader set of light curve data given in Table 7. 

   Once the observation times are converted to phase using the period or period function, we use a cubic spline to interpolate the light curves and find the magnitude at the observed phase, $m_\phi$.
The phase-correction curve $\mathcal{C}_\phi$ is defined as
\begin{equation*}
\mathcal{C}_\phi = \overline{m} - m_\phi,
\end{equation*}
where $\overline{m}$ is the mean magnitude.
  To transform the phase corrections from the ground-based system to the WFC3 system we use the color transformations from \citet{Riess:2016}:
\begin{eqnarray*}
m_{555} & = & V + 0.034 + 0.11 (V-I), \\
m_{814} & = &  I + 0.02 -0.018 (V-I), \\
m_{160} & = & H  + 0.16 (J-H).
\end{eqnarray*}
\noindent
We note that the WFC3 phase-correction curves are independent of the zeropoints of the ground-based light curves or the zeropoint transformation between the ground and WFC3, e.g.,
\begin{eqnarray*}
&\mathcal{C}_\phi(F555W) &= (1 + 0.11) \mathcal{C}_\phi(V) - 0.11 \mathcal{C}_\phi(I), \\
&\mathcal{C}_\phi(F814W) &= (1 + 0.018) \mathcal{C}_\phi(I) - 0.018 \mathcal{C}_\phi(V), \\
&\mathcal{C}_\phi(F160W) &= (1 - 0.16) \mathcal{C}_\phi(H) + 0.16 \mathcal{C}_\phi(J).
\end{eqnarray*}

 For the $H$ band, we fit the templates from \citet{Inno:2015} to the data if the number of $H$-band measurements is fewer than 20 (SS~CMa and XY~Car). In some cases there are no ground-based data ($I$-band data for S~Vul, $J$-band data for SS~CMa and XY~Car), so we predicted the ground phase-correction curve based on those of the two neighboring bands:
\begin{equation*}
\mathcal{C}_\phi(m_1) = a_0 + a_1\, \cos(2\pi\phi)+ a_2\, \sin(2\pi\phi) + a_3\, \mathcal{C}_\phi(m_2) + a_4\, \mathcal{C}_\phi(m_3).
\end{equation*}
\noindent
The coefficients $a_0$--$a_4$ were derived from least-squares regression over a sample of $\sim 50$ Milky Way Cepheids. For each {\it HST} measurement, we computed its phase $\Phi$ with $\Phi = (t \mod p) / p$, where $p = p_0$ or $p = p_0 + p_1\, t$, depending on which $\omega$ model is used for the corresponding Cepheid. The mean magnitude is then obtained by adding the phase-correction value $\mathcal{C}_{\phi=\Phi}$ to the {\it HST} measurement.

\begin{deluxetable}{ccccc}
\setlength\tabcolsep{0.4cm}
\tablecaption{Sources of the Ground-Based VIJH Light Curves\label{tbl_src}}
\tablehead{
\colhead{Identifier} & \multicolumn4c{References$^a$} \\
\cline{2-5}
& \colhead{$V$} & \colhead{$I$} & \colhead{$J$} & \colhead{$H$}}
\startdata
S Vul & 1, 2 & NA & 3 & 3 \\
SS CMa & 1, 4 & 5, 6 & NA & 7 \\
XY Car & 4 & 8 & NA & 7 \\
VX Per & 9--15 & 15 & 3 & 3 \\
VY Car & 5, 13, 16 & 5, 6 & 17 & 7, 17 \\
WZ Sgr & 1, 4 & 5 & 3, 17 & 3, 7, 17 \\
X Pup & 1, 4 & 6, 18 & 17 & 7, 17 \\
\enddata
\tablecomments{
$^a$The labels correspond to the following references:\\
1: Unreleased ASAS (K. Stanek, private communication).\\
2: Vila observation (Engle, private communication).\\
3: \citet{Monson+2011}.\\
4: ASAS \citet{Pojmanski1997}.\\
5: \citet{Coulson+1985}; from McMaster.\\
6: \citet{Berdnikov+2000}.\\
7: CTIO observations. \\
8: \citet{Coulson+1985b}; from McMaster.\\
9: \citet{Berdnikov1992}; from McMaster.\\
10: \citet{Szabados1981}; from McMaster.\\
11: \citet{Szabados1991}; from McMaster.\\
12: \citet{Berdnikov1987}; from McMaster.\\
13: \citet{Harris1980}; from McMaster.\\
14: \citet{Berdnikov+1993}; from McMaster.\\
15: \citet{Moffett+1984}; from McMaster.\\
16: \citet{Madore1975}; from McMaster.\\
17: \citet{Laney+1992}; from McMaster.\\
18: \citet{Berdnikov+1995}; from McMaster.\\
NA: No ground-based light curves, linear interpolation used.
}
\setlength\tabcolsep{6pt} % set back to default value
\end{deluxetable}

\begin{deluxetable}{cccccc}
\tabletypesize{\scriptsize}
\tablecaption{Sources of the Ground Light Curves for Period Determination\label{tbl_prule}}
\tablehead{
\colhead{Identifier} & \colhead{References$^a$} & \colhead{bands} & \colhead{$p_0$} & \colhead{$p_1$} & \colhead{$t_\mathrm{ref}$}
}
\startdata
     S~Vul & 1--17 & $U$,$B$,$V$,$R_J$,$H^b$ &   68.96595 &  2.3362e-04 &   2455786.6\\
    SS~CMa & 1,4,6,7,18,19 & $U$,$B$,$V$,$R_C$,$I_C$ &   12.35586 & -2.3166e-07 &   2449153.6\\
    XY~Car & 4,6,16,18,20 & $U$,$B$,$V$,$R_C$,$I_C$ &   12.43585 &  4.6595e-08 &   2448352.1\\
    VX~Per & 6,8,13,16,21--24 & $V$,$U$,$B$,$R_J$ &   10.88654 & -1.7926e-07 &   2445958.3\\
    VY~Car & 5,16--19,25 & $B$,$V$,$R_C$,$I_C$,$J_\mathrm{SAAO}$,$K_\mathrm{SAAO}$,$H$ &   18.89843 & -1.3034e-06 &   2449417.1\\
    WZ~Sgr & 3--7,12,16--19,24,25 & $U$,$B$,$V$,$R_C$,$I_C$,$J_\mathrm{SAAO}$,$K_\mathrm{SAAO}$,$R_J$,$I_J$,$J_\mathrm{CTIO}$,$K_\mathrm{CTIO}$,$H$ &   21.85119 &             NA &   2449469.1\\
     X~Pup & 4--7,16--18,24,26--28 & $B$,$V$,$J_\mathrm{SAAO}$,$K_\mathrm{SAAO}$,$V_W$,$B_W$,$U_W$,$L_W$,$H$ &   25.96749 &  4.7212e-07 &   2448872.4\\
\enddata
\tablecomments{$^a$ The labels correspond the following references:\\
1: Unreleased ASAS (K. Stanek, private communication)\\
2: RCT observations (S. Engle, private communication) \\
3: \citet{Monson+2011} \\
4: ASAS; \citet{Pojmanski1997}\\
5: CTIO observations\\
6: I-OMC; \citet{Alfonso-Garzon+2012}\\
7: \citet{1986PZ.....22..369B}; from McMaster\\
8: \citet{Berdnikov1987}; from McMaster\\
9: \citet{1992AandAT....2..107B}; from McMaster\\
10: \citet{1992AandAT....2....1B}; from McMaster\\
11: \citet{1992AandAT....2...31B}; from McMaster\\
12: \citet{1992AandAT....2...43B}; from McMaster\\
13: \citet{Berdnikov1992}; from McMaster\\
14: \citet{1993AstL...19...84B}; from McMaster\\
15: \citet{1995AstL...21..308B}; from McMaster\\
16: \citet{Harris1980}; from McMaster\\
17: \citet{Laney+1992}; from McMaster\\
18: \citet{Madore1975}; from McMaster\\
19: \citet{Coulson+1985}; from McMaster\\
20: \citet{Coulson+1985b}; from McMaster\\
21: \citet{Szabados1981}; from McMaster\\
22: \citet{Szabados1991}; from McMaster\\
23: \citet{Berdnikov+1993}; from McMaster\\
24: \citet{Moffett+1984}; from McMaster\\
25: \citet{Welch+1984}; from McMaster\\
26: \citet{Berdnikov+1995}; from McMaster\\
27: \citet{Pel1976}; from McMaster\\
28: \citet{Welch1986}; from McMaster\\
$^b$ Data merged from $H_\mathrm{SAAO}$ and $H_\mathrm{CTIO}$ measurements.
}
\end{deluxetable}

 \vfill
\eject

\input specpar.tex
\input obs.tex

\clearpage
\bibliographystyle{aasjournal} %
\bibliography{bibdesk}
\clearpage

\include{figs4}

\end{document}

%% file: specpar.tex
\startlongtable
\begin{deluxetable*}{ccccccccccccc}
\tabletypesize{\footnotesize}
\tablewidth{0pt}
\tablenum{2}
\tablecaption{Reference Stars\label{tb:h0var}}
\tablehead{\colhead{Star} &\colhead{R.A.}&\colhead{Decl. (deg)}&\colhead{Class}&\colhead{Quality}&\colhead{Source}&\colhead{$T_{\rm eff}$}&\colhead{$\sigma$}&\colhead{Log g}&\colhead{$\sigma$}&\colhead{F606W}&\colhead{$\mu$ (mag)}&\colhead{$\sigma$}}
\startdata
sscma0003 & 111.4971  & -25.2325  & K0IV & fair & Gemini & 5340  & 156  & 3.74  & 0.5  & 13.85  & 11.66  & 0.10  \\
sscma0005 & 111.5198  & -25.2406  & A7V & good & Gemini & 7920  & 254  & 3.89  & 0.5  & 12.34  & 9.779  & 0.11  \\
sscma0006 & 111.5137  & -25.2626  & G5IV & vgood & N/A & 5730  & 114  & 3.80  & 0.5  & 16.56  & 10.80  & 0.13  \\
sscma0010 & 111.5417  & -25.2306  & F7IV & vgood & Gemini & 6550  & 205  & 3.89  & 0.5  & 16.33  & 11.33  & 0.13  \\
sscma0014 & 111.5199  & -25.2360  & F5IV-V & vgood & Lick & 6700  & 134  & 3.96  & 0.5  & 12.33  & 9.142  & 0.12  \\
sscma0025 & 111.5430  & -25.2747  & G5IV-V & good & Lick & 5730  & 114  & 3.89  & 0.5  & 14.72  & 10.72  & 0.14  \\
sscma0026 & 111.5227  & -25.2806  & A2III-IV & vgood & Lick & 8675  & 362  & 4.44  & 0.5  & 13.71  & 11.57  & 0.14  \\
sscma0029 & 111.5334  & -25.2550  & G6V & vgood & Gemini & 5693  & 133  & 4.24  & 0.5  & 15.33  & 10.16  & 0.11  \\
sscma0031 & 111.5465  & -25.2718  & F8IV & vgood & Gemini & 6425  & 156  & 3.88  & 0.5  & 15.46  & 11.07  & 0.14  \\
sscma0038 & 111.5387  & -25.2663  & G0V & vgood & Lick & 5745  & 328  & 4.23  & 0.5  & 15.55  & 11.14  & 0.11  \\
sscma0039 & 111.5639  & -25.2815  & G0V & fair & Lick & 6300  & 180  & 4.12  & 0.5  & 16.69  & 11.34  & 0.15  \\
sscma0043 & 111.5217  & -25.2864  & K5III & good & Keck & 5500  & 188  & 1.00  & 0.1  & 15.26  & 17.67  & 0.15  \\
sscma0044 & 111.5229  & -25.2878  & G2IV-V & good & Lick & 5800  & 185  & 3.90  & 0.5  & 15.33  & 10.88  & 0.10  \\
sscma0045 & 111.5305  & -25.2872  & F8II-III & fair & Lick & 6300  & 128  & 3.20  & 0.6  & 15.58  & 13.09  & 0.10  \\
sscma0066 & 111.5503  & -25.2656  & G6IV-V & good & Lick & 5513  & 133  & 3.86  & 0.5  & 17.08  & 11.89  & 0.65  \\
sscman001 & 111.5079  & -25.2345  & NA & NA & NA & 0  & 0  & 0.00  & 0.0  & 0.000  & 11.10  & 0.17  \\
\hline
xycar0065 & 165.5151  & -64.2920  & M5III & poor & Gemini & 5440  & 133  & 0.00  & 0.0  & 16.59  & 13.85  & 0.10  \\
xycar0068 & 165.6194  & -64.2607  & F8IV-V & good & Gemini & 6300  & 156  & 3.94  & 0.5  & 15.99  & 11.23  & 0.15  \\
xycar0071 & 165.6018  & -64.2744  & G9IV & vgood & Gemini & 5390  & 156  & 3.75  & 0.5  & 15.81  & 11.80  & 0.25  \\
xycar0083 & 165.5281  & -64.2914  & K4 & fair & Gemini & 4600  & 260  & 0.00  & 0.0  & 15.44  & 13.10  & 0.15  \\
xycar0098 & 165.5259  & -64.2498  & F8V & vgood & Gemini & 6425  & 156  & 4.10  & 0.5  & 14.75  & 10.36  & 0.15  \\
xycar0100 & 165.5023  & -64.2788  & A1IV-V & vgood & Gemini & 9185  & 339  & 3.94  & 0.5  & 14.67  & 11.75  & 0.10  \\
xycar0105 & 165.5614  & -64.2864  & K4II & good & Gemini & 5020  & 509  & 1.95  & 0.7  & 12.51  & 12.01  & 0.22  \\
xycar0121 & 165.5442  & -64.2957  & M5III & poor & Gemini & 5800  & 185  & 0.00  & 0.0  & 15.28  & 10.45  & 0.52  \\
xycar0123 & 165.5456  & -64.2951  & G8IV & good & SOAR & 5440  & 133  & 3.76  & 0.5  & 15.60  & 11.58  & 0.19  \\
xycar0179 & 165.5647  & -64.2606  & A4IV & good & Gemini & 8200  & 339  & 3.98  & 0.5  & 14.29  & 11.05  & 0.13  \\
xycar0243 & 165.5724  & -64.2306  & A6IV & good & Gemini & 8212  & 254  & 3.99  & 0.5  & 16.39  & 12.41  & 0.10  \\
\hline
vxper0030 & 31.98145  & 58.42410  & G0IV-V & good & Keck & 5925  & 180  & 3.91  & 0.5  & 16.12  & 10.76  & 0.60  \\
vxper0035 & 32.00800  & 58.42643  & F4V & good & Lick & 6662  & 134  & 4.06  & 0.5  & 15.47  & 11.01  & 0.14  \\
vxper0036 & 31.90272  & 58.46210  & kA6hA8mF2-- & NA & Keck & 7920  & 254  & 0.00  & 0.0  & 15.34  & 11.54  & 0.10  \\
vxper0039 & 31.99930  & 58.43603  & F1IV & vgood & Keck & 7275  & 205  & 3.95  & 0.5  & 15.00  & 10.84  & 0.10  \\
vxper0041 & 31.98719  & 58.45766  & mA6V & NA & Lick & 7730  & 248  & 0.00  & 0.0  & 14.61  & 11.39  & 0.12  \\
vxper0042 & 31.91492  & 58.43107  & F9IV-V & vgood & Keck & 6237  & 180  & 3.93  & 0.5  & 14.11  & 9.639  & 0.20  \\
vxper0043 & 31.92551  & 58.46995  & G5V & vgood & Lick & 5730  & 114  & 4.23  & 0.5  & 14.10  & 8.968  & 0.31  \\
vxper0044 & 31.93525  & 58.46506  & F8III & good & Hydra & 6175  & 156  & 3.42  & 0.6  & 14.43  & 10.10  & 0.76  \\
vxper0049 & 31.92722  & 58.43344  & B1V & vgood & Keck & 26100  & 1000  & 3.67  & 0.5  & 11.36  & 12.08  & 0.11  \\
vxper0067 & 31.88844  & 58.44349  & K5I & good & Lick & 4600  & 260  & 0.97  & 0.7  & 11.77  & 12.09  & 0.18  \\
vxpern001 & 31.93254  & 58.46487  & F5V & fair & Lick & 6700  & 116  & 4.05  & 0.5  & 16.49  & 11.77  & 0.10  \\
\hline
wzsgr0090 & 274.2439  & -19.0481  & G5II & fair & Keck & 5230  & 153  & 2.13  & 0.7  & 15.40  & 12.97  & 0.10  \\
wzsgr0109 & 274.2284  & -19.0566  & K0III-IV & fair & Lick & 4920  & 156  & 2.88  & 0.7  & 14.45  & 12.88  & 0.10  \\
wzsgr0119 & 274.2306  & -19.0513  & G9II & fair & Lick & 5390  & 156  & 2.27  & 0.7  & 14.39  & 12.77  & 0.10  \\
wzsgr0120 & 274.2613  & -19.0543  & K4III & good & Lick & 4600  & 188  & 2.42  & 0.7  & 14.04  & 11.62  & 0.11  \\
wzsgr0122 & 274.2547  & -19.0919  & G7IV-V & vgood & Lick & 5656  & 133  & 3.88  & 0.5  & 14.75  & 9.446  & 0.37  \\
wzsgr0124 & 274.2527  & -19.0568  & F9V & vgood & Lick & 6237  & 180  & 4.13  & 0.5  & 14.00  & 9.609  & 0.12  \\
wzsgr0162 & 274.2926  & -19.0923  & F8V & good & Lick & 6425  & 156  & 4.10  & 0.5  & 15.55  & 11.40  & 0.13  \\
wzsgr0191 & 274.2534  & -19.0908  & G9V & fair & Lick & 5070  & 156  & 4.39  & 0.5  & 16.36  & 12.08  & 0.10  \\
wzsgr0269 & 274.2708  & -19.0476  & K0III-IV & good & Lick & 5340  & 156  & 3.25  & 0.6  & 13.53  & 11.75  & 0.13  \\
wzsgr0275 & 274.2674  & -19.0473  & NA & NA & NA & 0  & 0  & 0.00  & 0.0  & 16.01  & 11.97  & 0.11  \\
wzsgr0306 & 274.2657  & -19.0323  & G3V & good & Lick & 6026  & 153  & 4.17  & 0.5  & 16.81  & 11.18  & 0.12  \\
wzsgr0310 & 274.2623  & -19.0389  & F4III & fair & Lick & 7347  & 212  & 3.79  & 0.5  & 15.66  & 11.29  & 0.12  \\
wzsgrn003 & 274.2803  & -19.0822  & NA & NA & NA & 0  & 0  & 0.00  & 0.0  & 12.67  & 11.11  & 0.28  \\
\hline
svul0008 & 297.1140  & 27.30387  & G6II & good & Lick & 5513  & 133  & 2.38  & 0.7  & 13.81  & 13.01  & 0.10  \\
svul0015 & 297.0976  & 27.31197  & F0V & good & Lick & 7635  & 248  & 3.93  & 0.5  & 14.78  & 11.72  & 0.19  \\
svul0016 & 297.0758  & 27.28951  & G7III & fair & Lick & 5156  & 208  & 2.87  & 0.7  & 15.13  & 12.86  & 0.13  \\
svul0019 & 297.1244  & 27.28699  & F6V & good & Lick & 6550  & 205  & 4.08  & 0.5  & 15.21  & 11.15  & 0.32  \\
svul0020 & 297.1009  & 27.26104  & K6-- & good & Keck & 5440  & 133  & 0.00  & 0.0  & 15.70  & 14.17  & 0.10  \\
svul0021 & 297.0982  & 27.27661  & G1V & fair & Lick & 5865  & 175  & 4.20  & 0.5  & 15.71  & 10.39  & 0.49  \\
svul0028 & 297.1217  & 27.28104  & F2II & good & Lick & 7200  & 205  & 4.03  & 0.5  & 16.08  & 13.03  & 0.10  \\
svul0041 & 297.1052  & 27.29179  & F1V & good & Keck & 7275  & 205  & 3.97  & 0.5  & 15.83  & 12.86  & 0.11  \\
svul0042 & 297.1041  & 27.24918  & G7III-IV & good & Keck & 5476  & 133  & 3.35  & 0.6  & 16.21  & 13.32  & 0.10  \\
svul0062 & 297.0741  & 27.31131  & G9IV & good & Keck & 5390  & 156  & 3.75  & 0.5  & 16.68  & 13.19  & 0.11  \\
svul0074 & 297.1284  & 27.28546  & F5Ia & fair & Lick & 7310  & 212  & 1.88  & 0.7  & 17.01  & 13.14  & 0.10  \\
svuln009 & 297.0882  & 27.31561  & K6-- & poor & Lick & 5440  & 133  & 0.00  & 0.0  & 16.89  & 14.17  & 0.10  \\
svuln016 & 297.0818  & 27.29010  & G8III & good & Lick & 5440  & 133  & 3.05  & 0.6  & 15.63  & 13.39  & 0.10  \\
\hline
xpup0004 & 113.1950  & -20.9228  & B7V & fair & Keck & 13500  & 647  & 3.59  & 0.5  & 12.87  & 12.88  & 0.10  \\
xpup0008 & 113.1849  & -20.8820  & NA & NA & NA & 5440  & 133  & 0.00  & 0.0  & 14.43  & 13.03  & 0.10  \\
xpup0011 & 113.2106  & -20.9232  & F2V & vgood & Hydra & 7200  & 205  & 3.98  & 0.5  & 14.96  & 11.38  & 0.10  \\
xpup0014 & 113.2090  & -20.8689  & F5V & good & Lick & 6875  & 134  & 4.03  & 0.5  & 15.02  & 11.14  & 0.11  \\
xpup0017 & 113.2245  & -20.9229  & A0V & vgood & Hydra & 9600  & 1000  & 3.75  & 0.5  & 15.04  & 12.11  & 0.15  \\
xpup0021 & 113.2240  & -20.8950  & F8V & vgood & Keck & 6300  & 156  & 4.12  & 0.5  & 15.63  & 11.16  & 0.11  \\
xpup0027 & 113.2173  & -20.8660  & F7V & vgood & Lick & 6350  & 205  & 4.11  & 0.5  & 0.000  & 11.22  & 0.12  \\
xpup0028 & 113.1836  & -20.8876  & F4V & good & Keck & 7907  & 212  & 3.89  & 0.5  & 16.00  & 11.91  & 0.10  \\
xpup0032 & 113.2124  & -20.8985  & F9IV & good & Keck & 6362  & 180  & 3.88  & 0.5  & 16.40  & 11.65  & 0.12  \\
xpup0036 & 113.2067  & -20.8719  & F1V & good & Keck & 7125  & 205  & 3.99  & 0.5  & 16.58  & 12.28  & 0.10  \\
xpup0044 & 113.2284  & -20.8957  & F2IV-V & vgood & Keck & 6800  & 205  & 3.96  & 0.5  & 16.90  & 12.19  & 0.12  \\
xpup0048 & 113.2042  & -20.8753  & G9III-IV & good & Keck & 5390  & 156  & 3.28  & 0.6  & 17.00  & 13.66  & 0.10  \\
xpup0050 & 113.1978  & -20.8819  & G9V & good & Keck & 5390  & 156  & 4.31  & 0.5  & 17.44  & 11.89  & 0.34  \\
xpupn000 & 113.2137  & -20.9025  & NA & NA & NA & 0  & 0  & 0.00  & 0.0  & 17.33  & 9.881  & 0.28  \\
xpupn001 & 113.2109  & -20.9313  & NA & NA & NA & 0  & 0  & 0.00  & 0.0  & 17.86  & 11.73  & 0.10  \\
xpupn002 & 113.2162  & -20.9311  & NA & NA & NA & 0  & 0  & 0.00  & 0.0  & 17.52  & 9.881  & 0.28  \\
xpupn004 & 113.2137  & -20.9025  & NA & NA & NA & 0  & 0  & 0.00  & 0.0  & 17.33  & 13.43  & 0.35  \\
\hline
vycar0010 & 161.1465  & -57.5821  & F2V & good & AAT & 7200  & 205  & 3.98  & 0.5  & 13.69  & 11.11  & 0.23  \\
vycar0021 & 161.1578  & -57.5941  & A0V & vgood & Gemini & 9600  & 1000  & 3.75  & 0.5  & 14.74  & 12.30  & 0.17  \\
vycar0029 & 161.0890  & -57.5733  & G0V & good & N/A & 6175  & 180  & 4.14  & 0.5  & 15.77  & 11.09  & 0.16  \\
vycar0051 & 161.1569  & -57.5693  & G9IV-V & vgood & Gemini & 5390  & 156  & 3.85  & 0.5  & 13.39  & 8.944  & 5.49  \\
vycar0060 & 161.1321  & -57.5448  & mAV4 & NA & SOAR & 7920  & 254  & 0.00  & 0.0  & 12.72  & 10.35  & 0.26  \\
vycar0077 & 161.1772  & -57.5360  & K5III-IV & vgood & N/A & 4680  & 260  & 2.64  & 0.7  & 15.10  & 13.31  & 0.10  \\
vycar0078 & 161.1733  & -57.5367  & F1III-IV & vgood & Gemini & 7275  & 205  & 4.19  & 0.5  & 14.90  & 11.36  & 0.12  \\
vycar0091 & 161.1576  & -57.5454  & F1III-IV & good & Gemini & 7275  & 205  & 4.19  & 0.5  & 15.86  & 12.11  & 0.19  \\
vycar0109 & 161.1333  & -57.5529  & F0V & vgood & Gemini & 7350  & 248  & 3.96  & 0.5  & 15.80  & 12.13  & 0.12  \\
vycar0124 & 161.1082  & -57.5855  & F4IV-V & good & Gemini & 7062  & 134  & 3.97  & 0.5  & 15.40  & 12.72  & 0.12  \\
vycar0129 & 161.1388  & -57.5943  & kA1hA7mA9-- & NA & Gemini & 7920  & 254  & 0.00  & 0.0  & 14.80  & 12.23  & 0.13  \\
vycar0147 & 161.0889  & -57.5815  & F5IV-V & vgood & AAT & 6875  & 134  & 3.97  & 0.5  & 13.46  & 10.26  & 0.43  \\
vycar0148 & 161.1004  & -57.5866  & NA & NA & NA & 7920  & 254  & 0.00  & 0.0  & 16.98  & 13.18  & 0.10  \\
vycar0150 & 161.0962  & -57.5918  & A7 & fair & Gemini & 6300  & 156  & 0.00  & 0.0  & 16.50  & 11.79  & 0.11  \\
vycar0153 & 161.0987  & -57.5874  & F0 & fair & Gemini & 6300  & 156  & 0.00  & 0.0  & 17.26  & 12.51  & 0.12  \\
vycar0179 & 161.0943  & -57.5729  & K3V & good & Gemini & 5120  & 412  & 4.38  & 0.5  & 16.22  & 9.762  & 0.15  \\
vycar0226 & 161.1098  & -57.5324  & K3III & vgood & AAT & 5120  & 412  & 2.84  & 0.7  & 11.91  & 11.12  & 0.43  \\
vycar0231 & 161.1043  & -57.5396  & NA & NA & NA & 0  & 0  & 0.00  & 0.0  & 13.41  & 11.63  & 0.20  \\
\hline
\enddata
\tablecomments{blah}
\end{deluxetable*}

%% file: obs.tex
\startlongtable
\begin{deluxetable*}{ccccccccc}
\tabletypesize{\footnotesize}
\tablewidth{0pt}
\tablenum{1}
\tablecaption{Spatial Scanning Observations Used in This Paper\label{tb:h0var}}
\tablehead{\colhead{Epoch} &\colhead{Date}&\colhead{Program ID }&\colhead{Scan Rates}&\colhead{Filters}&\colhead{Scan Lengths}&\colhead{PA V3 (deg)}&\colhead{X pos targ}&\colhead{File root}}
\startdata
\hline
 \multicolumn{9}{l}{sscma} \\ 
\hline
1 &  2012-10-23 & 12879 & 0.330 0.330  & F606W F621M & 115.5 115.5  & 116.00  & -3.10 -2.99  & ibzc04 \\
1 &  2012-10-23 & 12879 & 1.505  & F606W & 526.7  & 116.00  & -2.14  & ibzc04 \\
2 &  2013-04-18 & 12879 & 0.330 0.330  & F606W F621M & 114.8 114.8  & 295.99  & -3.10 -2.99  & ibzc15 \\
2 &  2013-04-18 & 12879 & 1.505  & F606W & 523.7  & 295.99  & -2.13  & ibzc15 \\
3 &  2013-10-22 & 13344 & 0.330 0.330  & F606W F621M & 114.8 114.8  & 116.00  & -3.10 -2.99  & ic8z04 \\
3 &  2013-10-22 & 13344 & 1.505  & F606W & 523.7  & 116.00  & -2.13  & ic8z04 \\
4 &  2014-04-16 & 13344 & 0.330 0.330  & F606W F621M & 114.8 114.8  & 295.99  & 2.898 3.000  & ic8z15 \\
4 &  2014-04-16 & 13344 & 1.505  & F606W & 523.7  & 295.99  & 9.860  & ic8z15 \\
5 &  2014-10-23 & 13678 & 0.330 0.330  & F606W F621M & 114.8 114.8  & 116.00  & -3.10 -3.10  & icir03 \\
7 &  2015-10-23 & 14206 & 0.330 0.330  & F606W F621M & 114.8 114.8  & 116.00  & -3.10 -2.99  & ictn06 \\
7 &  2015-10-23 & 14206 & 1.505  & F606W & 523.7  & 116.00  & -2.13  & ictn06 \\
8 &  2016-04-15 & 14206 & 0.330 0.330  & F606W F621M & 114.8 114.8  & 295.99  & 2.898 3.000  & ictn07 \\
8 &  2016-04-15 & 14206 & 1.505  & F606W & 523.7  & 295.99  & 17.86  & ictn07 \\
9 &  2016-10-22 & 14648 & 0.330 0.330  & F606W F621M & 114.8 114.8  & 116.00  & -3.10 -3.10  & id5d06 \\
9 &  2016-10-22 & 14648 & 0.330  & F606W & 114.8  & 115.99  & -2.99  & id5d06 \\
a &  2017-04-14 & 14648 & 0.330 0.330  & F606W F621M & 114.8 114.8  & 295.99  & 2.898 3.000  & id5d11 \\
a &  2017-04-14 & 14648 & 1.505  & F606W & 523.7  & 295.99  & 17.86  & id5d11 \\
\hline
 \multicolumn{9}{l}{xycar} \\ 
\hline
1 &  2012-08-07 & 12879 & 0.410 0.410  & F606W F621M & 143.5 143.5  & 337.01  & -3.12 -2.99  & ibzc02 \\
2 &  2013-01-26 & 12879 & 0.410 0.430  & F606W F621M & 143.5 149.6  & 156.98  & -3.12 -2.99  & ibzc11 \\
3 &  2013-08-08 & 13344 & 0.410 0.410  & F606W F621M & 143.5 143.5  & 337.01  & -3.12 -2.99  & ic8z02 \\
4 &  2014-01-25 & 13344 & 0.410 0.430  & F606W F621M & 142.6 149.6  & 156.98  & 2.873 3.001  & ic8z11 \\
5 &  2014-08-07 & 13678 & 0.410 0.410  & F606W F621M & 143.5 143.5  & 337.01  & -3.12 -3.12  & icir01 \\
7 &  2015-08-09 & 14206 & 0.410 0.410  & F606W F621M & 143.5 143.5  & 337.01  & -3.12 -2.99  & ictn01 \\
8 &  2016-01-27 & 14206 & 0.410 0.410  & F606W F621M & 143.5 143.5  & 156.98  & 2.872 3.001  & ictn02 \\
9 &  2016-08-09 & 14648 & 0.410 0.410  & F606W F621M & 143.5 143.5  & 337.01  & -3.12 -3.12  & id5d01 \\
a &  2017-01-22 & 14648 & 0.410 0.410  & F606W F621M & 143.5 143.5  & 156.98  & 2.872 3.001  & id5d22 \\
\hline
 \multicolumn{9}{l}{vxper} \\ 
\hline
1 &  2013-02-07 & 12879 & 0.410 0.430  & F606W F621M & 143.5 150.5  & 250.02  & -3.12 -2.99  & ibzc13 \\
2 &  2013-08-28 & 12879 & 0.410 0.430  & F606W F621M & 143.5 150.5  & 69.971  & -3.12 -2.99  & ibzc23 \\
3 &  2014-02-09 & 13344 & 0.410 0.430  & F606W F621M & 143.5 150.5  & 250.02  & -3.12 -2.99  & ic8z13 \\
4 &  2014-08-17 & 13344 & 0.410 0.430  & F606W F621M & 142.6 149.6  & 69.970  & 2.873 3.001  & ic8z23 \\
5 &  2015-02-09 & 13678 & 0.410 0.430  & F606W F621M & 143.5 150.5  & 250.02  & -3.12 -3.13  & icir11 \\
6 &  2015-08-17 & 14206 & 0.410 0.430  & F606W F621M & 142.6 149.6  & 69.970  & 2.873 3.001  & ictn05 \\
7 &  2016-02-07 & 14206 & 0.410 0.430  & F606W F621M & 142.6 149.6  & 250.02  & -3.12 -2.99  & ictn08 \\
8 &  2016-08-27 & 14648 & 0.410 0.430  & F606W F621M & 142.6 149.6  & 69.970  & 2.873 2.867  & id5d05 \\
9 &  2017-02-07 & 14648 & 0.410 0.430  & F606W F621M & 142.6 149.6  & 250.02  & -3.12 -2.99  & id5d23 \\
\hline
 \multicolumn{9}{l}{wzsgr} \\ 
\hline
1 &  2013-09-24 & 13334 & 0.410 0.430  & F606W F673N & 143.5 149.6  & 288.99  & -3.12 -3.13  & ic8k13 \\
2 &  2014-03-25 & 13334 & 0.410 0.430  & F606W F673N & 143.5 149.6  & 109.00  & -3.12 -3.13  & ic8k14 \\
3 &  2014-09-23 & 13686 & 0.410 0.430  & F606W F673N & 143.5 149.6  & 288.99  & -3.12 -3.13  & iciu13 \\
4 &  2015-03-25 & 13686 & 0.410 0.430  & F606W F673N & 143.5 149.6  & 109.00  & -3.12 -3.13  & iciu23 \\
5 &  2015-09-23 & 14062 & 0.410 0.430  & F606W F673N & 143.5 149.6  & 288.99  & -3.12 -3.13  & ict704 \\
6 &  2016-03-25 & 14206 & 0.410 0.430  & F606W F673N & 143.5 149.6  & 109.00  & -3.12 -3.13  & ictn14 \\
7 &  2016-09-24 & 14206 & 0.410 0.430  & F606W F673N & 143.5 149.6  & 288.99  & -3.12 -3.13  & ictn13 \\
8 &  2017-03-25 & 14648 & 0.410 0.430  & F606W F673N & 143.5 149.6  & 109.00  & -3.12 -3.13  & id5d14 \\
\hline
 \multicolumn{9}{l}{svul} \\ 
\hline
1 &  2013-10-29 & 13334 & 0.410 0.690  & F606W F621M & 143.5 241.5  & 277.00  & -3.12 -0.99  & ic8k03 \\
2 &  2014-03-28 & 13334 & 0.410 0.690  & F606W F621M & 143.5 241.5  & 96.993  & 2.872 5.002  & ic8k04 \\
3 &  2014-10-26 & 13686 & 0.410 0.690  & F606W F621M & 143.5 241.5  & 277.00  & -3.12 -0.99  & iciu17 \\
4 &  2015-03-30 & 13686 & 0.410 0.690  & F606W F621M & 143.5 241.5  & 96.993  & 2.872 5.002  & iciu18 \\
5 &  2015-10-29 & 14062 & 0.410 0.690  & F606W F621M & 143.5 241.5  & 277.00  & -3.12 -0.99  & ict706 \\
6 &  2016-03-28 & 14206 & 0.410 0.690  & F606W F621M & 143.5 241.5  & 96.993  & 2.872 10.00  & ictn16 \\
7 &  2016-10-25 & 14206 & 0.410 0.690  & F606W F621M & 143.5 241.5  & 277.00  & -3.12 -0.99  & ictn15 \\
8 &  2017-03-28 & 14648 & 0.410 0.690  & F606W F621M & 143.5 241.5  & 96.993  & 2.872 10.00  & id5d15 \\
\hline
 \multicolumn{9}{l}{xpup} \\ 
\hline
1 &  2013-10-24 & 13334 & 0.410 0.350  & F606W F673N & 143.5 129.5  & 122.00  & -3.12 -3.11  & ic8k11 \\
2 &  2014-04-16 & 13334 & 0.410 0.350  & F606W F673N & 143.5 129.5  & 301.99  & -3.12 -3.11  & ic8k12 \\
3 &  2014-10-24 & 13686 & 0.410 0.350  & F606W F673N & 143.5 129.5  & 122.00  & -3.12 -3.11  & iciu21 \\
4 &  2015-04-18 & 13686 & 0.410 0.350  & F606W F673N & 143.5 129.5  & 301.99  & -3.12 -3.11  & iciu22 \\
5 &  2015-10-24 & 14062 & 0.410 0.350  & F606W F673N & 143.5 129.5  & 122.00  & -3.12 -3.11  & ict707 \\
6 &  2016-04-16 & 14206 & 0.410 0.350  & F606W F673N & 143.5 129.5  & 301.99  & -3.12 -3.11  & ictn18 \\
7 &  2016-10-23 & 14206 & 0.410 0.350  & F606W F673N & 143.5 129.5  & 122.00  & -3.12 -3.11  & ictn17 \\
8 &  2017-04-16 & 14648 & 0.410 0.350  & F606W F673N & 143.5 129.5  & 301.99  & -3.12 -3.11  & id5d16 \\
\hline
 \multicolumn{9}{l}{vycar} \\ 
\hline
1 &  2013-07-09 & 13334 & 0.410 0.860  & F606W F673N & 143.5 301.0  & 327.00  & -3.12 -0.99  & ic8k07 \\
2 &  2014-01-10 & 13334 & 0.410 0.860  & F606W F673N & 143.5 301.0  & 146.99  & 2.872 5.003  & ic8k08 \\
3 &  2014-07-28 & 13686 & 0.410 0.860  & F606W F673N & 143.5 301.0  & 327.00  & -3.12 -0.99  & iciu07 \\
4 &  2015-01-12 & 13686 & 0.410 0.860  & F606W F673N & 143.5 301.0  & 146.99  & 2.872 5.003  & iciu20 \\
5 &  2015-07-07 & 14062 & 0.410 0.860  & F606W F673N & 143.5 301.0  & 327.00  & -3.12 -0.99  & ict703 \\
6 &  2016-01-10 & 14206 & 0.410 0.860  & F606W F673N & 143.5 301.0  & 146.99  & 2.872 9.003  & ictn12 \\
7 &  2016-07-07 & 14206 & 0.410 0.860  & F606W F673N & 143.5 301.0  & 327.00  & -3.12 -0.99  & ictn19 \\
8 &  2017-01-12 & 14648 & 0.410 0.860  & F606W F673N & 143.5 301.0  & 146.99  & 2.872 9.003  & id5d13 \\
\enddata
\tablecomments{There are 2 exposures for each unique entry except the 1.5"/sec scans for sscma.  All exposure times are 348-350 seconds.   }
\end{deluxetable*}

%% file: figs4.tex
\begin{figure}[ht]
\vspace*{220mm}
\figurenum{1}
\includegraphics{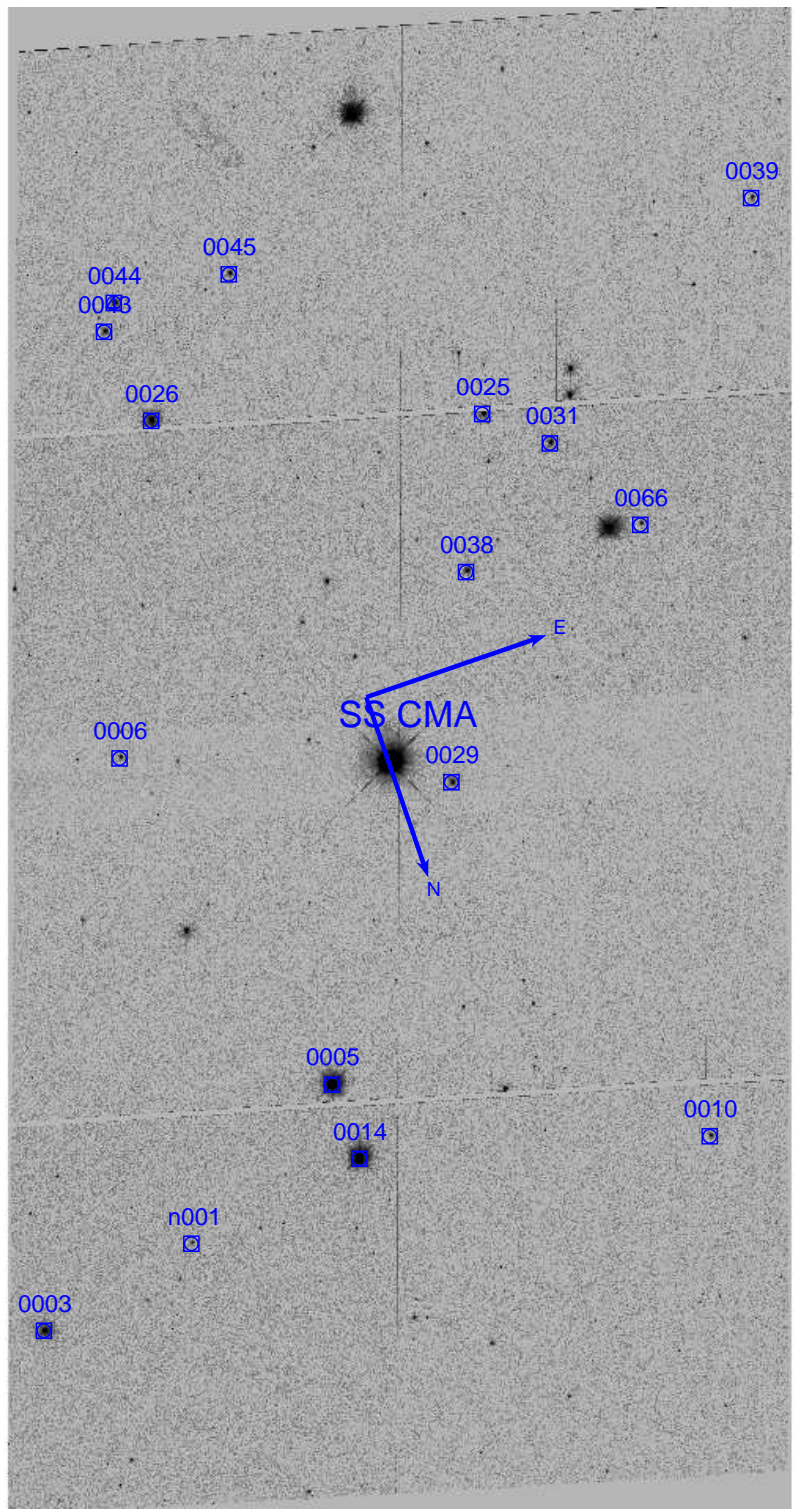}
\caption {{\it HST} WFC3-UVIS images ($2.7' \times 4.7'$) of the field centered around
Cepheid SS CMA covered by WFC3-UVIS spatial scanning.  Compass indicates direction of north and east.  Spectrophotometric reference stars are labeled.}
\end{figure}

\vfill \eject

\begin{figure}[ht]
\vspace*{220mm}
\figurenum{2}
\includegraphics{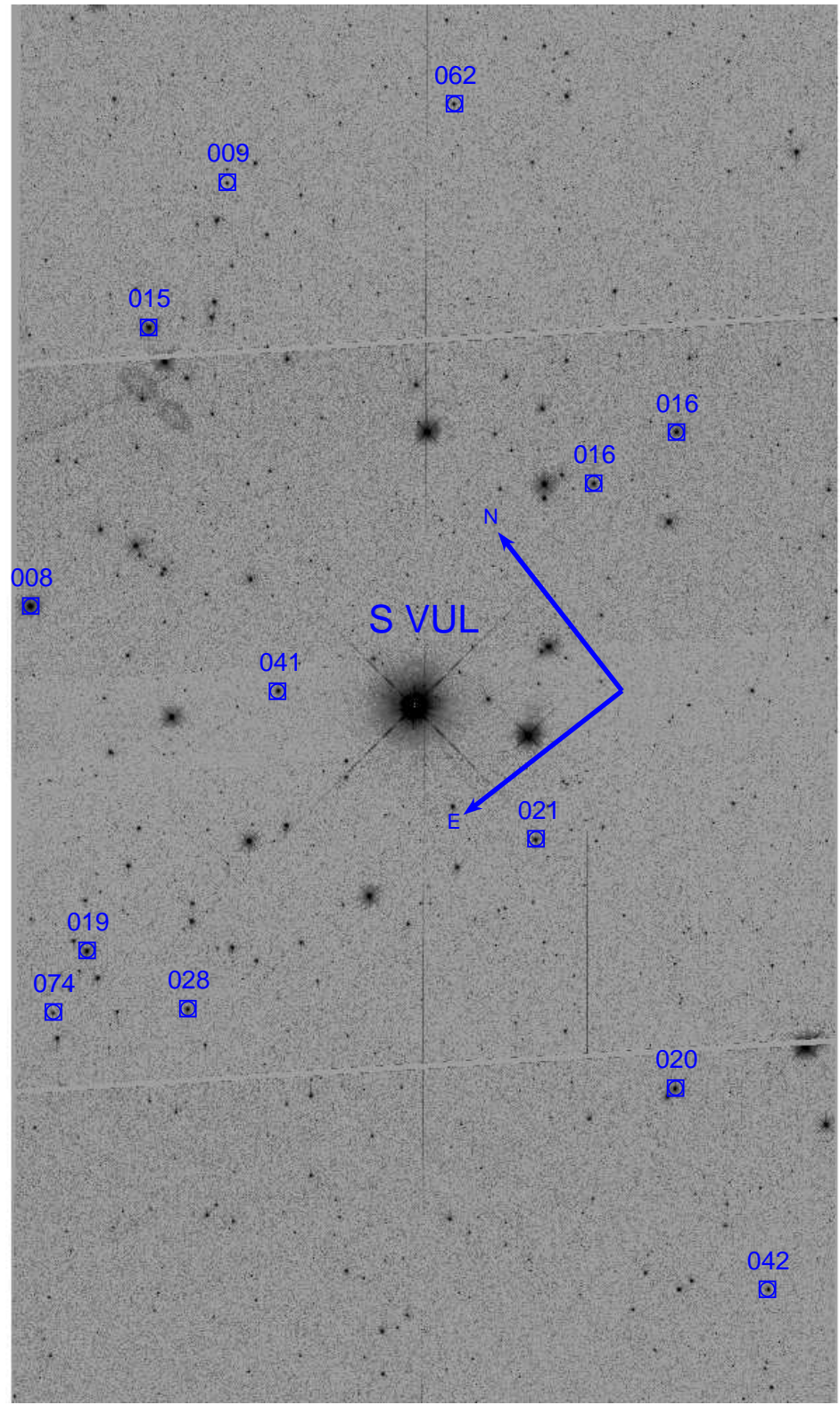}
\caption {Same as Figure 1, for S Vul.}
\end{figure}

\vfill \eject

\begin{figure}[ht]
\vspace*{220mm}
\figurenum{3}
\includegraphics{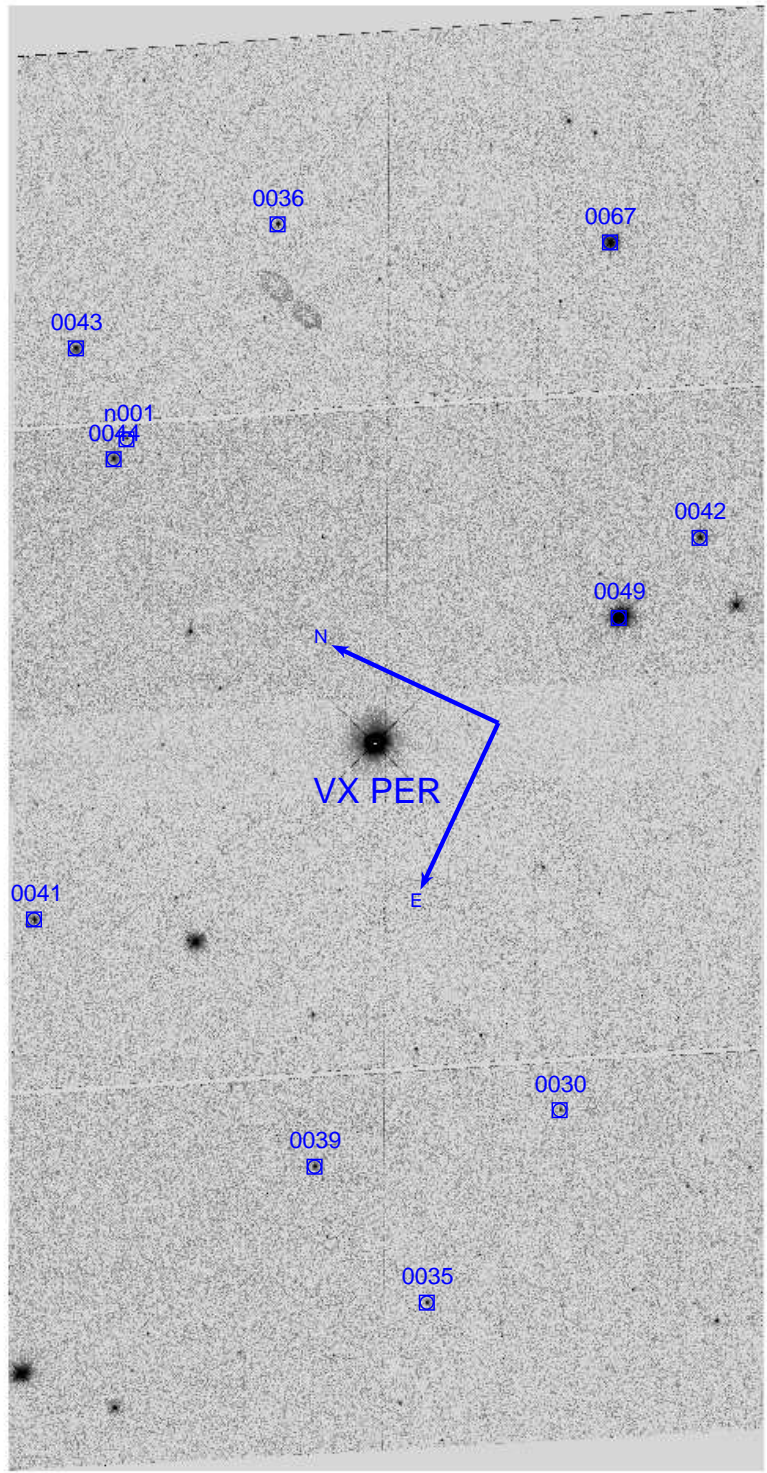}
\caption {Same as Figure 1, for VX Per.}
\end{figure}

\vfill \eject

\begin{figure}[ht]
\vspace*{220mm}
\figurenum{4}
\includegraphics{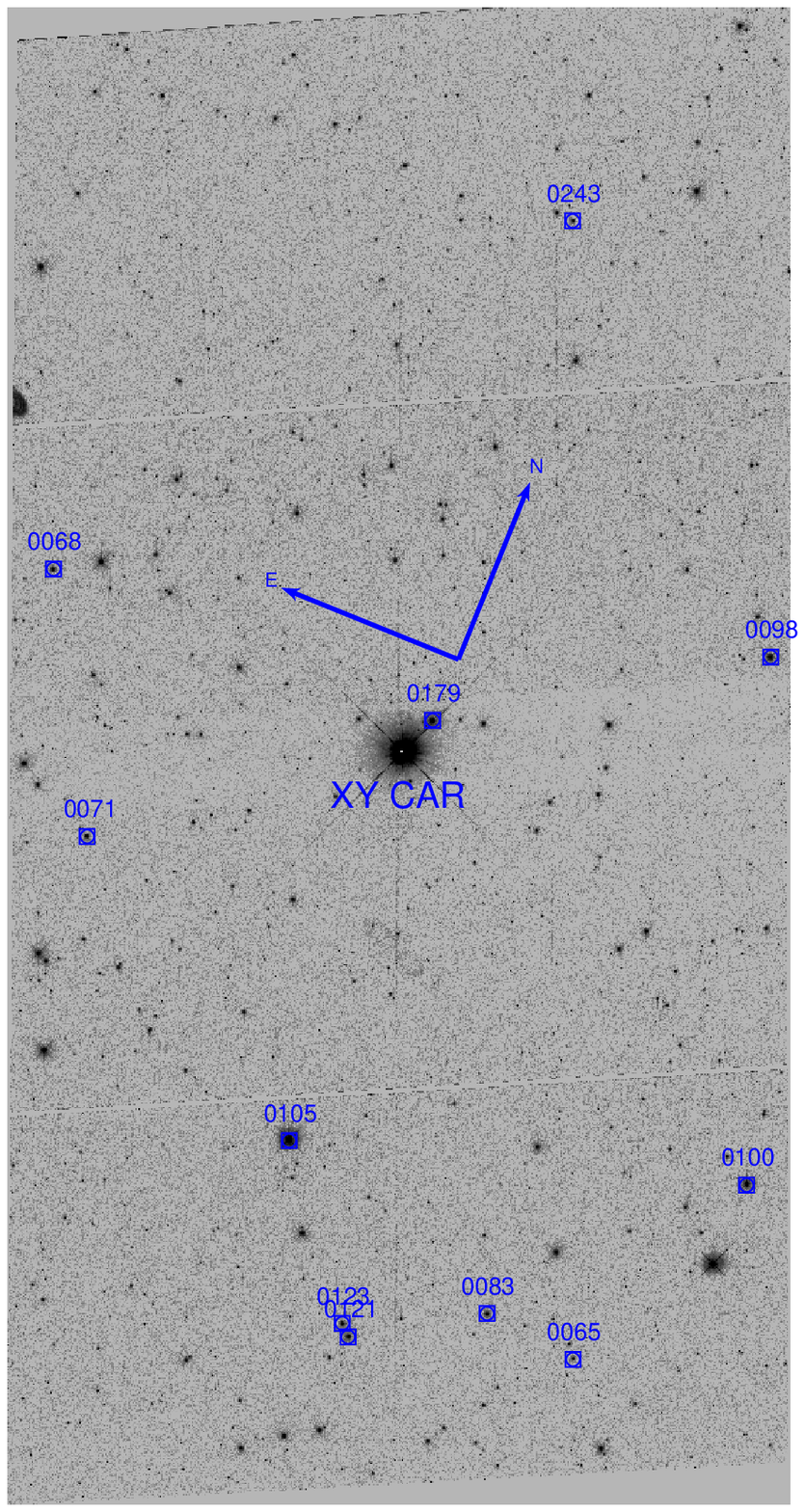}
\caption {Same as Figure 1, for XY Car.}
\end{figure}

\vfill \eject

\begin{figure}[ht]
\vspace*{220mm}
\figurenum{5}
\includegraphics{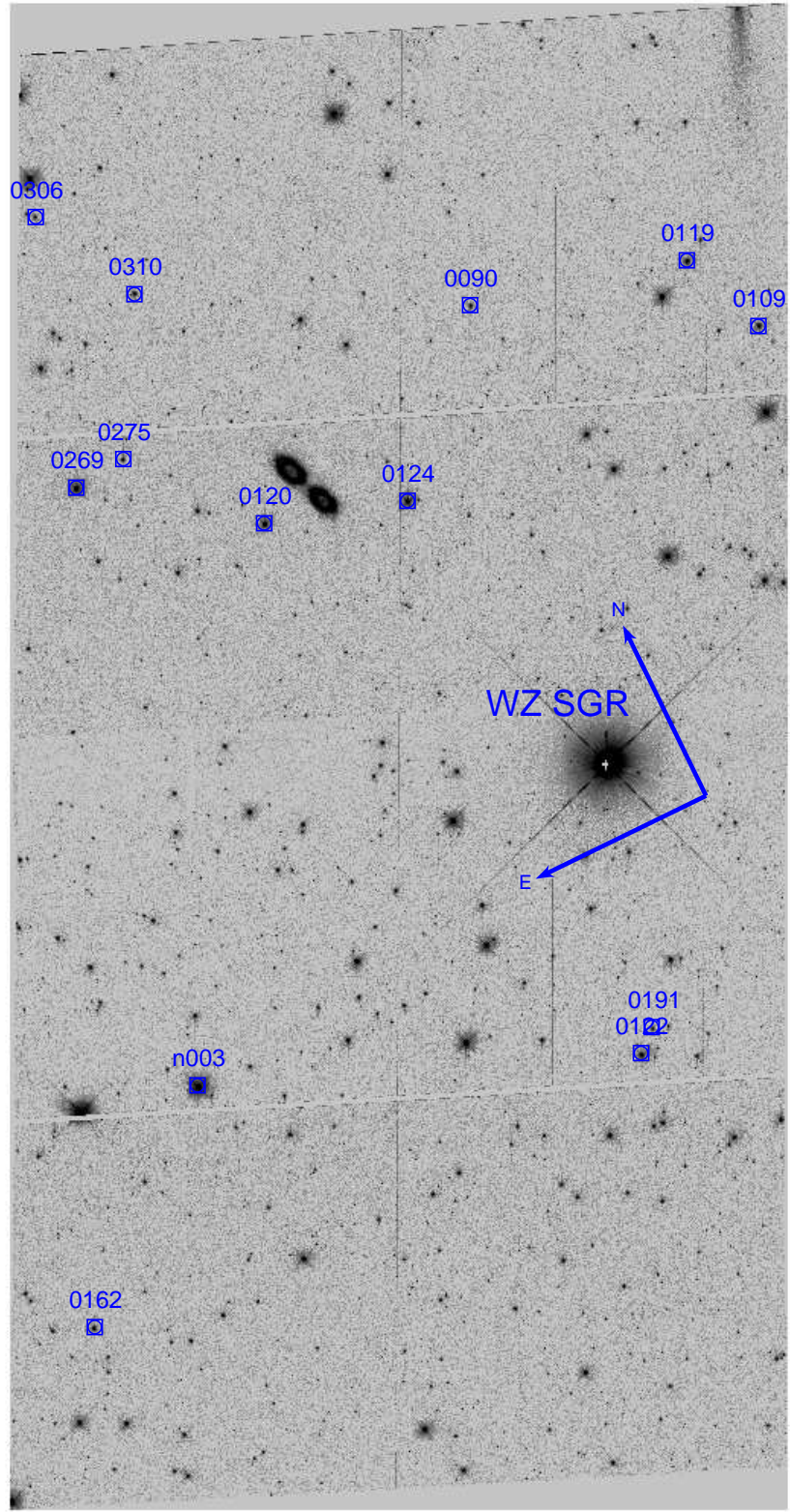}
\caption {Same as Figure 1, for WZ Sgr.}
\end{figure}

\vfill \eject

\begin{figure}[ht]
\vspace*{220mm}
\figurenum{6}
\includegraphics{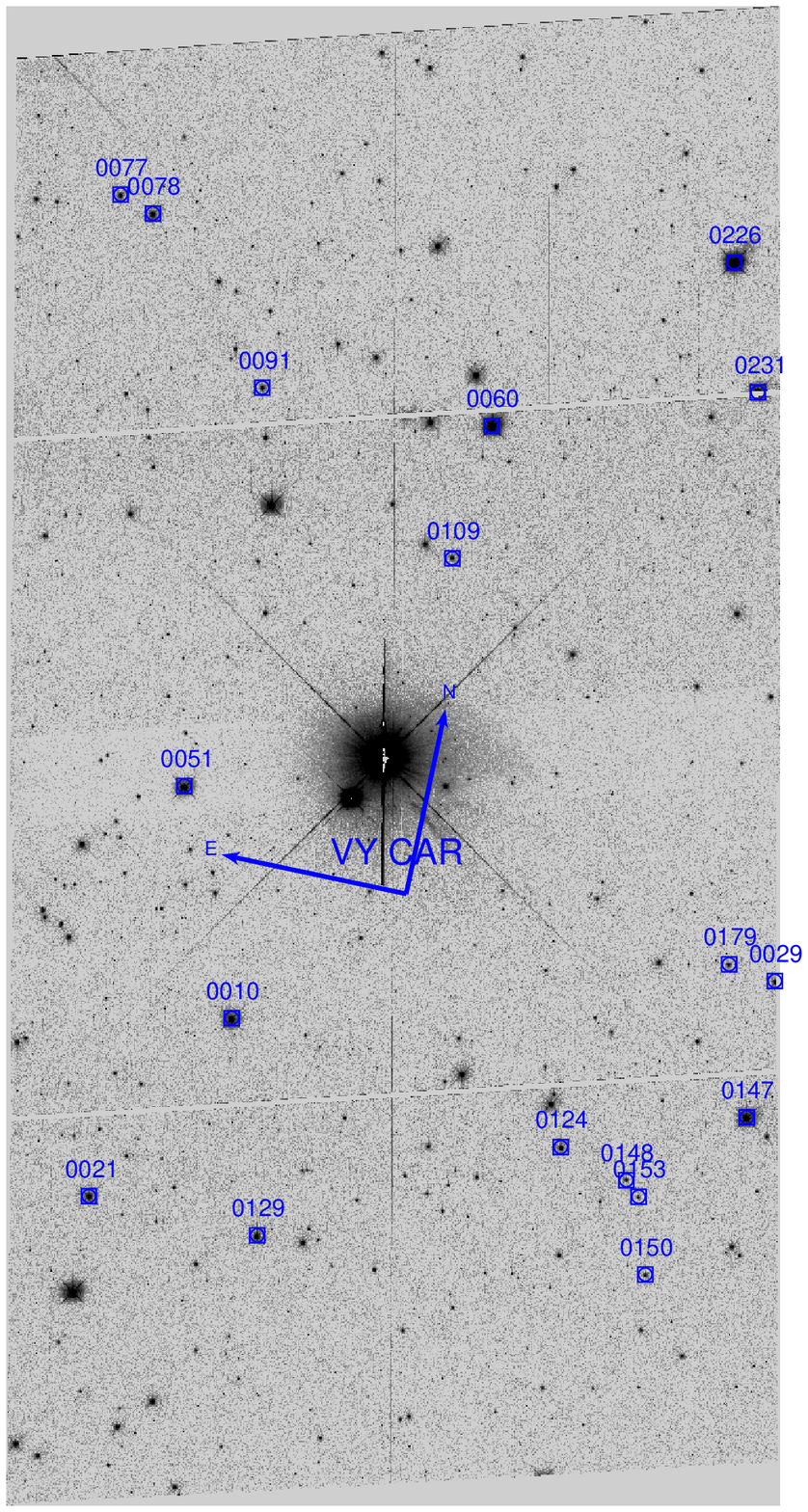}
\caption {Same as Figure 1, for VY Car.}
\end{figure}

\vfill \eject

\begin{figure}[ht]
\vspace*{220mm}
\figurenum{7}
\includegraphics{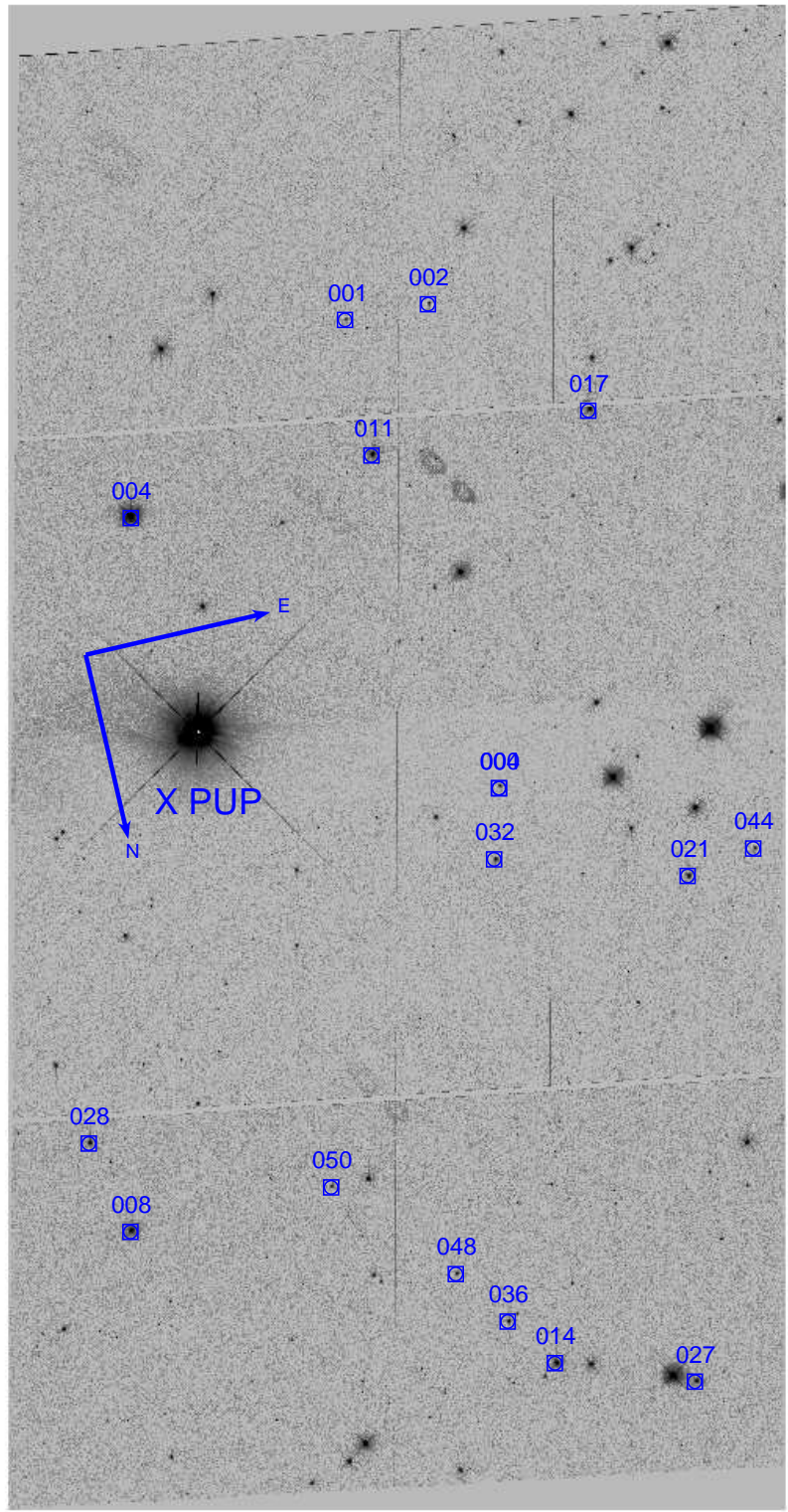}
\caption {Same as Figure 1, for X Pup.}
\end{figure}

\vfill \eject

\begin{figure}[ht]
\vspace*{150mm}
\figurenum{8}
\includegraphics{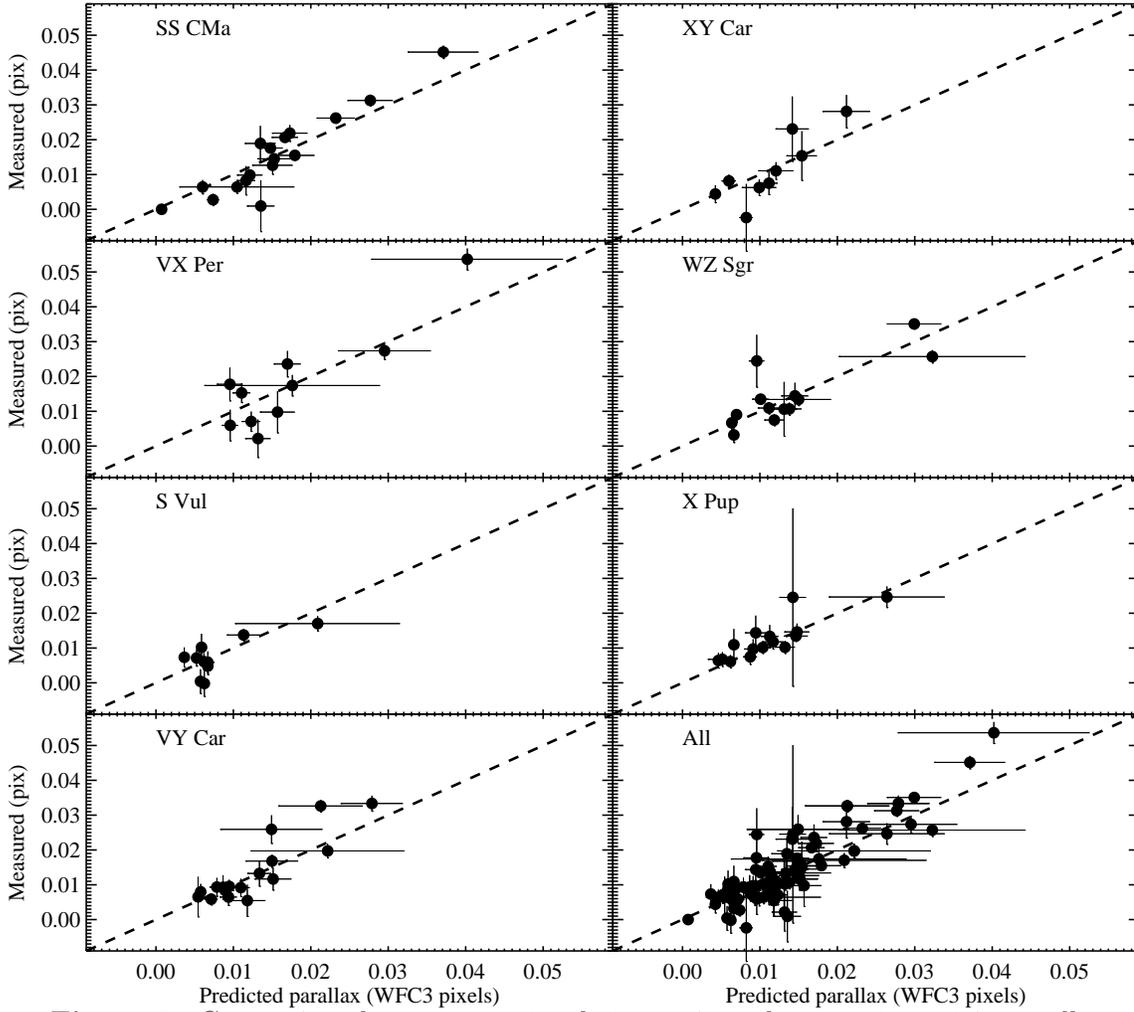}
\caption{\label{fg:plr} Comparison between spectrophotometric and pure astrometric parallax, obtained by excluding the spectrophotometric prior for each star in turn.  In this comparison the astrometric parallax is fully independent of the spectrophotometric parallax.}
\end{figure}

\begin{figure}[ht]
\vspace*{150mm}
\figurenum{9}
\includegraphics{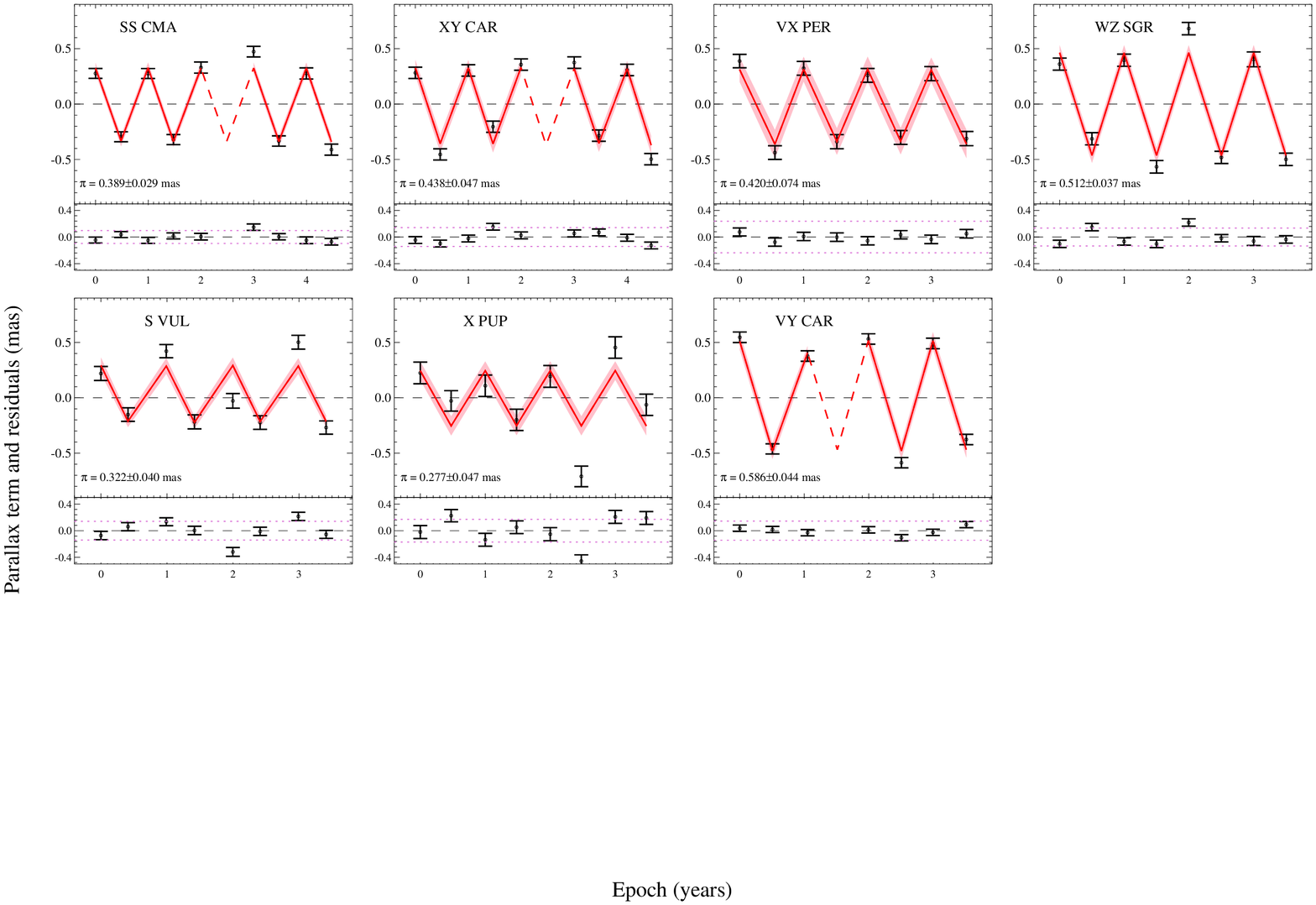}
\caption{\label{fg:plr} The proper-motion-subtracted 1D motion of the Cepheids observed at 6 month epochs spaced over 4~yr.  The red line indicates the best model fit of the parallax motion.  Plotted values include the parallax factor.}
\end{figure}

\begin{figure}[ht]
\vspace*{150mm}
\figurenum{10}
\includegraphics{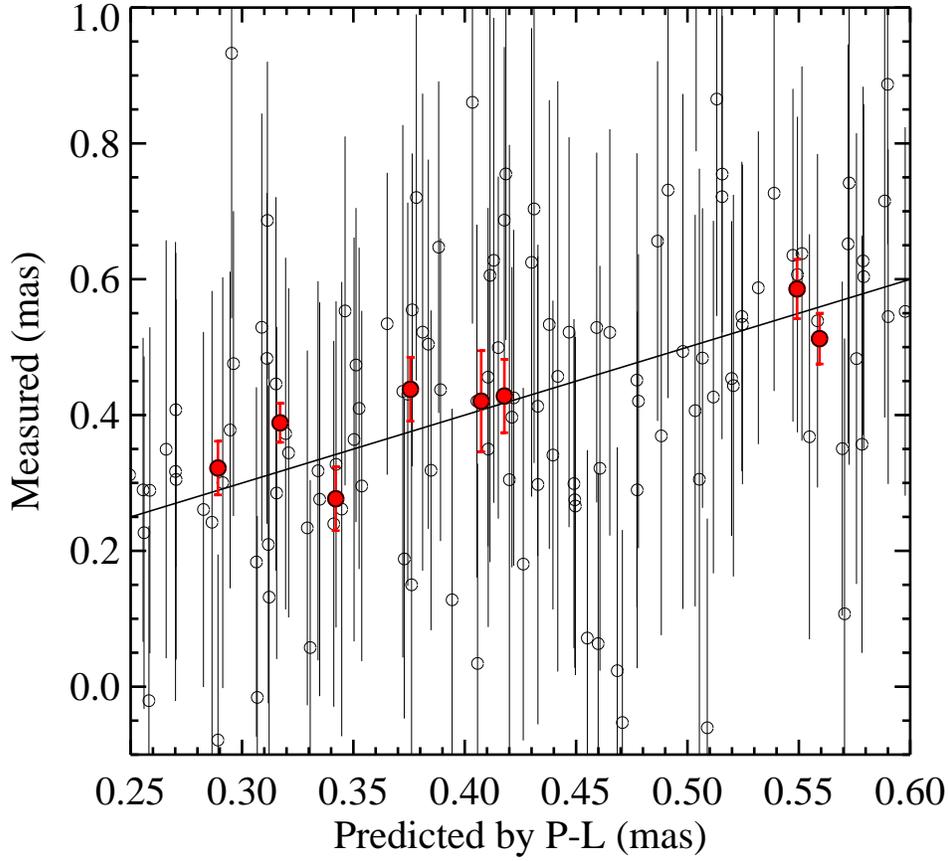}
\caption{\label{fg:parvpar} Comparison of predicted and measured parallaxes.  The predictions are based on their Wesenheit apparent magnitudes and periods as well as the Cepheid $P$--$L$ calibrated in R16 which yields H$_0=73.2$ km s$^{-1}$ Mpc$^{-1}$.  The 8 points in red are from the spatial scanning program presented here.  The open symbols are based on {\it Gaia} DR1.}
\end{figure}

\begin{figure}[ht]
\vspace*{150mm}
\figurenum{11}
\includegraphics{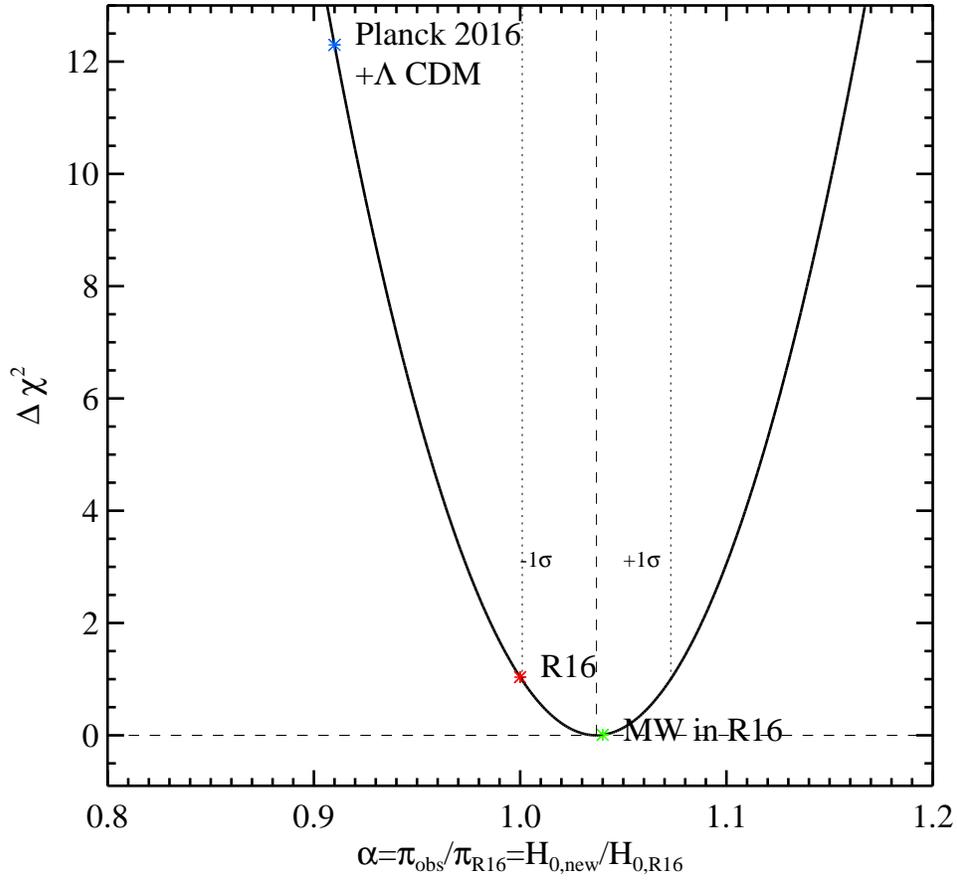}
\caption{\label{fg:chi} Change in $\chi^2$ resulting from comparing the measured parallaxes of 8 MW Cepheids to the values predicted by reversing the distance ladder (i.e., using the Hubble constant in R16), that is $\pi_{obs}=\alpha \pi_{R16}$ and $H_{0,new}=\alpha (H_{0,R16})$.  Fractions less than unity indicate a lower Hubble constant.  Position of the {\it Planck} 2016 + $\Lambda$CDM and the R16 result using only MW Cepheids as an anchor are indicated.}
\end{figure}

\begin{figure}[ht]
\vspace*{200mm}
\figurenum{12}
\includegraphics{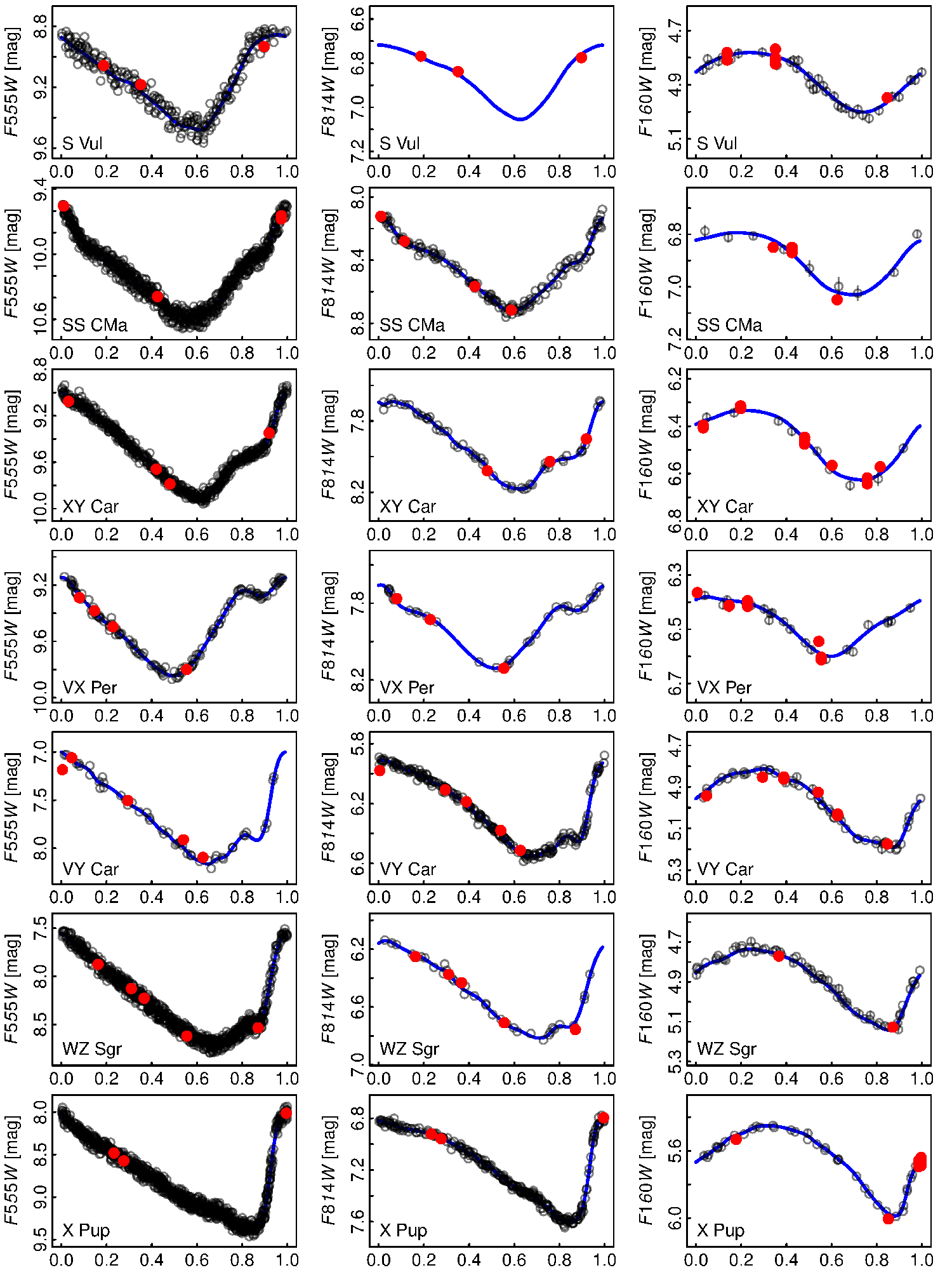}
\caption {Use of ground-based light curves (transformed to the WFC3 system) to determine the phase correction for the {\it HST} observations (red points) in $F555W$, $F814W$, and $F160W$.}
\end{figure}

\begin{figure}[ht]
\vspace*{150mm}
\figurenum{13}
\includegraphics{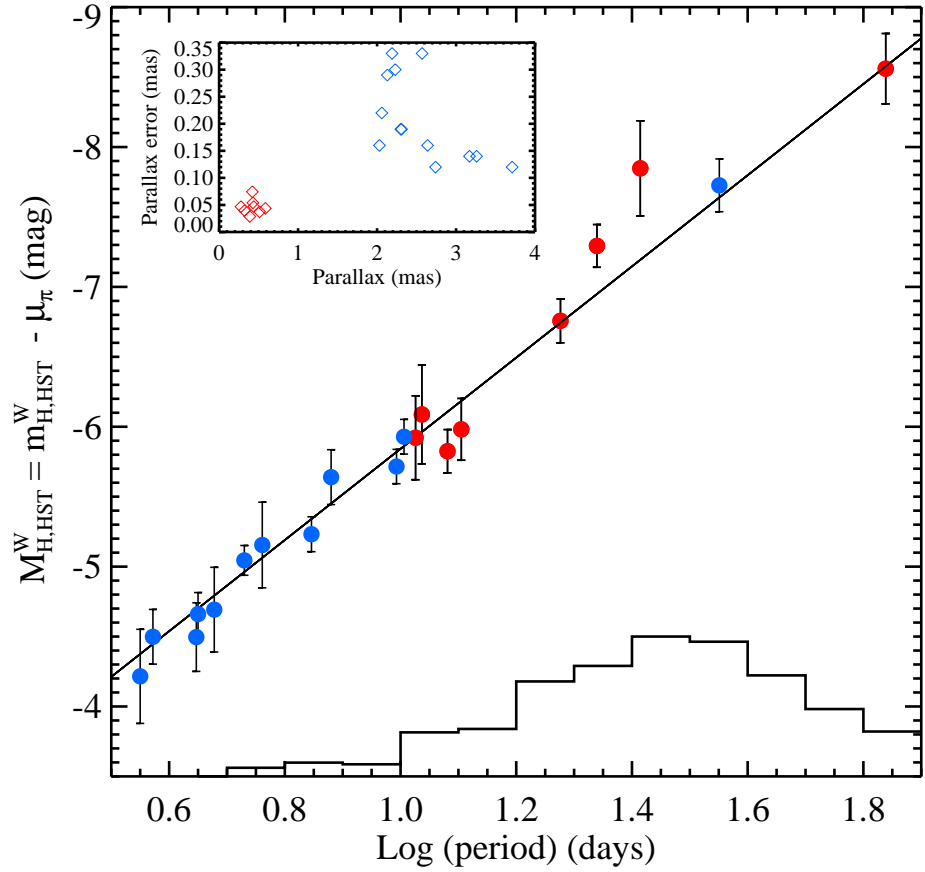}
\caption{The $P$--$L$ relation of Milky Way Cepheids based on trigonometric parallax measurements.  The points in blue were measured with the {\it HST} FGS \citep{benedict07} and {\it Hipparcos} \citep{vanleeuwen07} and are all within 0.5 kpc, and the points in red are presented here from spatial scanning of WFC3 and are in the range of $1.7 < D < 3.6$ kpc.  The inset shows the uncertainties in the measured parallaxes.}
\end{figure}